\documentclass[onecolumn,showpacs,preprintnumbers,amsmath,amssymb]{revtex4-1}
\usepackage[colorlinks=true,citecolor=blue,urlcolor=blue,linkcolor=blue,bookmarksopen]{hyperref}

\usepackage{titlesec}
\usepackage[pdftex]{graphicx}
\usepackage[utf8]{inputenc}
\usepackage[T1]{fontenc}
\usepackage{ulem}
\usepackage{mathrsfs}
\usepackage{graphicx}
\usepackage{natbib}
\usepackage{epsfig}
\usepackage{enumitem}
\usepackage{subcaption}

\begin{document}

	\title{Light Management in Perovskite Photovoltaic Solar Cells: a perspective}

	\author{Florian Berry$^1$}
	\author{Raphaël Mermet-Lyaudoz$^1$}
	\author{Jose Maria Cuevas Davila$^1$}
	\author{Djihad Amina Djemmah$^1$}
	\author{Hai Son Nguyen$^{1,2}$}
	\email{hai_son.nguyen@ec-lyon.fr}
	\author{Christian Seassal$^1$}
	\author{Erwann Fourmond$^1$}
	\author{Céline Chevalier$^1$}
	\author{Mohamed Amara$^1$}
	\author{Emmanuel Drouard $^1$}
	\email{emmanuel.drouard@ec-lyon.fr}

	\affiliation{$^1$Univ Lyon, Ecole Centrale de Lyon, CNRS, INSA Lyon, Université Claude Bernard Lyon 1, CPE Lyon, INL, UMR5270, 69130 Ecully, France} 
	\affiliation{$^2$Institut Universitaire de France (IUF)}

	\date{\today}

	\begin{abstract}
		Light Management (LM) is essential for metal-halide perovskite solar cells in their race for record performance. In this review, criteria on materials, processes and photonic engineering are established such as to enhance mainly the short circuit current density, towards high energy yields. These criteria are used to analyse a large panel of solutions envisaged in the literature for single junction cells. Moreover, a perspective based on rigorous electromagnetic simulations performed on various comparable structures is proposed in order to clarify the conclusions, and to pave the way to further performance enhancement in the case of all-perovskite, two-terminal tandem cells.\\

		KYEWORDS: light trapping, single junction perovskite cell, all perovskite tandem cells
	\end{abstract}
	\maketitle

	\section{Introduction}
	Perovskite photovoltaic solar cells (PPSC) have reached efficiency comparable to those of the best crystalline silicon (c-Si) devices within only a few years \cite{nrel.gov}, thanks to the tremendous electrical and optical properties of the metal-halide hybrid perovskites materials \cite{Snaith2013,jeon2015,Tress2017}. For single junction PPSC, the best performance yet demonstrated reach or even overpass a short-circuit current density $J_{sc}$ of 25.8 $mA.cm^{-2}$, an open circuit voltage $V_{oc}$ of 1.179 $V$, and a fill factor $FF$ of 0.846, leading to power conversion efficiency $PCE$ of 25.7\% \cite{nrel.gov}. These achievements are already quite close to the theoretically expected maximum values, especially for the $J_{sc}$, which is typically of 27.83 $mA.cm^{-2}$ under AM1.5G illumination, given that the band gap energy is estimated in their studies to 1.53 $eV$. Furthermore, tandem PPSC made of at least two different perovskite materials appears promising bust also challenging and less mature. In the case of the most common 2 terminal (2T) configuration, a net performance increase could have been recently demonstrated compared to single junction PPSC \cite{lin_all-perovskite_2022}. 
	
\begin{figure}[h]
		\centering
		\includegraphics[scale=0.6]{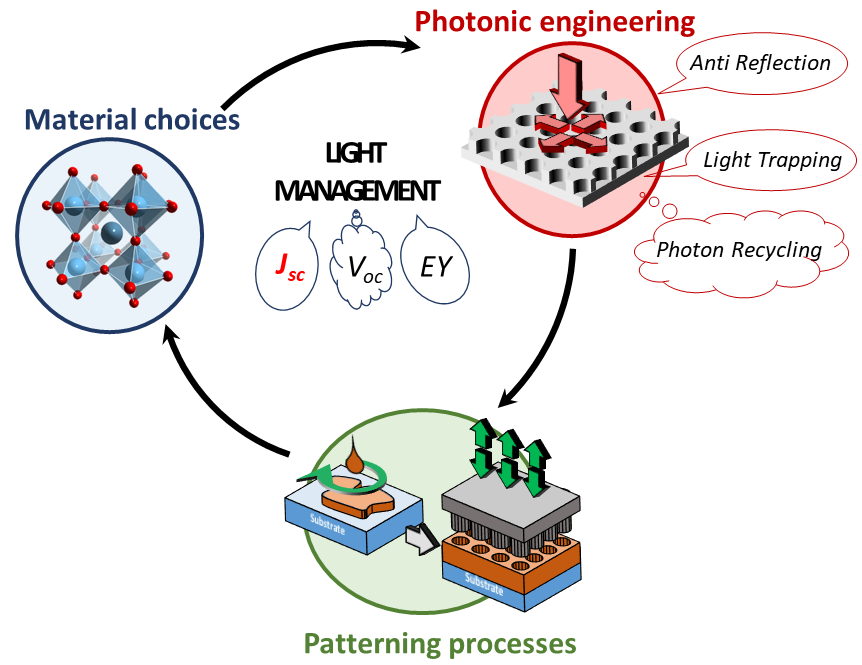}
		\caption{The three main correlated aspects that governed Light Management.}
		\label{fig:Interplays}
	\end{figure}

	As for every optoelectronic device, light management (LM) in PPSC is of primary importance to reach the best performance. As schematized in Figure \ref{fig:Interplays}, the LM is expected to impact three key performance parameters. First, thanks to the LM, impinging sunlight can be efficiently collected and then absorbed leading to a high $J_{sc}$. Then the LM can help to enhance the $V_{oc}$ thanks to the so-called photon recycling (PR)\cite{miller2012strong}, provided a strong external luminescence \cite{Ross1967} and considering the semiconductor exhibits weak non radiative recombinations. Finally, as recently proposed, the energy yield (EY) tends to substitute to the two previous criteria usually considered under standard test conditions (STC). For the latter, requested studies, however, remain complex since they intend to couple illumination models and angular-dependant optical, thermal and electrical models of the cell for one year \cite{Peters2018}. However, simply considering the incidence angle in the evaluation of absorption and thus $J_{sc}$ is a first step towards EY improvement.
	
	As illustrated in Figure \ref{fig:Interplays}, the LM in PPSC results from three interdependent considerations: the photonic engineering (PE) at the wavelength scale, the material choices, and the nanofabrication processes. Practically, the last two directly define the PPSC architecture.
	
	In brief, still regardless of the envisaged semiconductor, the LM can invoke several kinds of PE. Most frequently, it results in anti-reflection (AR) over the whole absorbed spectral range, leading to a larger $J_{sc}$ as well as possibly larger EY if this the AR effect also occurs over a wide angular domain. Moreover, beyond the AR effects, light trapping (LT) mechanisms can be involved at the band edge of the perovskite material, also leading to a larger $J_{sc}$ \cite{Bermel2007,Zhou2008}, whereas enhanced PR thanks to absorption suppression just above the band gap could enhance $V_{oc}$ \cite{Sandhu2013}, still provided previously mentioned conditions. This well established PE is reminded in more details in section \ref{light_management}. the LM could even be extended to the infrared domain to possibly limit the parasitic sub band gap absorption and enhance the thermal radiation, so as to decrease the operating temperature \cite{dumoulin_radiative_2021}. Finally, for the specific case of 2T tandem PPSC, the LM is required to ensure at the same time absorption optimization and current matching between the two cells. 
	
	In the specific case of metal-halide perovskites, intrinsic properties make these materials particularly relevant for LT, compared to the standard case of c-Si solar cells. First, metal-halide perovskites are direct band gap semiconductors, the light is then more easily absorbed in a submicron thick layer, where LT and possibly PR can occur at their band edge, as detailed later. Second, their refractive indices are lower than that of c-Si, thus, AR effect is easier to obtain over the spectral range of interest. Lastly, the ability to recycle photons has been observed/shown in the specific case of lead iodide perovskite material \cite{Pazos-Outon2016}.
	
	Other key properties of metal-halide perovskites, mainly electrical and with regards to aging, have direct consequences on the typical architecture of the PPSC. The diffusion lengths of the carriers in the perovskites imply both contacts should be in the vicinity of the photocarrier generation, thereby covering both sides of the cell {\cite{yang2019enhancing}}. Additionally, one of these contacts should obviously be transparent to the sunlight. The double heterojunction architecture is widespread, leading to the use of additional intermediate layers, the electron/hole transport layers (ETL/HTL). Then, the back metallic contact, ideally made of noble metal, can thus act as a back mirror. In any case, a careful optical design is required to keep low parasitic absorption in these surrounding layers. Moreover, an effective encapsulation should be added to prevent from the degradation of the perovskite. It should be carefully chosen considering the impact of its optical properties. 
	
	Let us finally envisage the wide panel of low-cost elaboration techniques and strategies enabling the structuration of the perovskite thin film at various scales, from the sub-micron to the over-micron one. First PPSC mainly relied on a mesoporous architecture, in which the perovskite infiltrated a scaffold. Then, progress in the elaboration techniques of the perovskite in liquid or vapor phases basically results in a better microcrystallization, and thus larger diffusion lengths. Simple planar architectures have therefore replaced mesoporous ones \cite{Liu2013,Liu2014b,Zhao2019}; thanks to these far larger diffusion lengths, current planar PPSC overpass the performances of the best mesoporous PPSC. 
	
	Following these huge modifications, the LM has evolved: if the mesoporous architecture could have been optimized for efficient LM thanks to the scaffold (typically made of TiO$_2$) to enhance the $J_{sc}$, most of the planar cells demonstrated in the last few years (see e.g. supporting information of ref \cite{Xu2020}) exhibit performances approaching the limits, up to about 90\% of the maximum achievable $J_{sc}$, mostly by simply taking care of AR.
	
	However, refined the LM still enables additional improvements of the performances towards the limits. For example, reduced optical losses thanks to LT especially at the band edge helps to gain the last missing $mA/cm^2$, as detailed in the following. In addition, the $V_{oc}$ could be improved thanks to an accurate treatment of the photocarrier recombination mechanisms and the associated rates.
	
	A few review papers discussed on various demonstrations of LM for PPSC at the time of their writing \cite{Zhang2018,Zhang2019}. In this paper, we focus on PE mainly in corrugated dielectric, without any additional metallic material inside or at the vicinity of the perovskite material. Indeed, if using metal can induce LT due to plasmonics effect, other phenomena such parasitic absorption can frequently balance this effect \cite{SiavashMoakhar2020}.
	
	After a brief presentation of the main PPSC architectures and material choices that impact LM, and summarizing the PE, we propose to further analyze various published structures relying on a wave optics approach. Thus, by unraveling the various mechanisms involved, we are able to compare the various proposed strategies to deduce guidelines for further optimization. These guidelines are finally illustrated thanks to simulations of the optical properties of a few case studies.
	
	\section{Key PPSC materials and architectures}
	
	The LM in the PPSC directly derive from their architecture, i.e. a stack of layers of various materials and thus optical indices, each layer having a thickness of the order of a few tenth of wavelengths, and with possible structurations from small to large scales compared to the wavelength. The main properties of the perovskite materials impacting LM have already been discussed. However, comparisons will appear difficult and need to be treated carefully, due to discrepancies between the structures investigated, as detailed hereafter. 
	
	\subsection{Patterning of the PPSC}
	\label{possible_architecture}
	
	\begin{figure}[h]
		\centering
		\includegraphics[width=\linewidth]{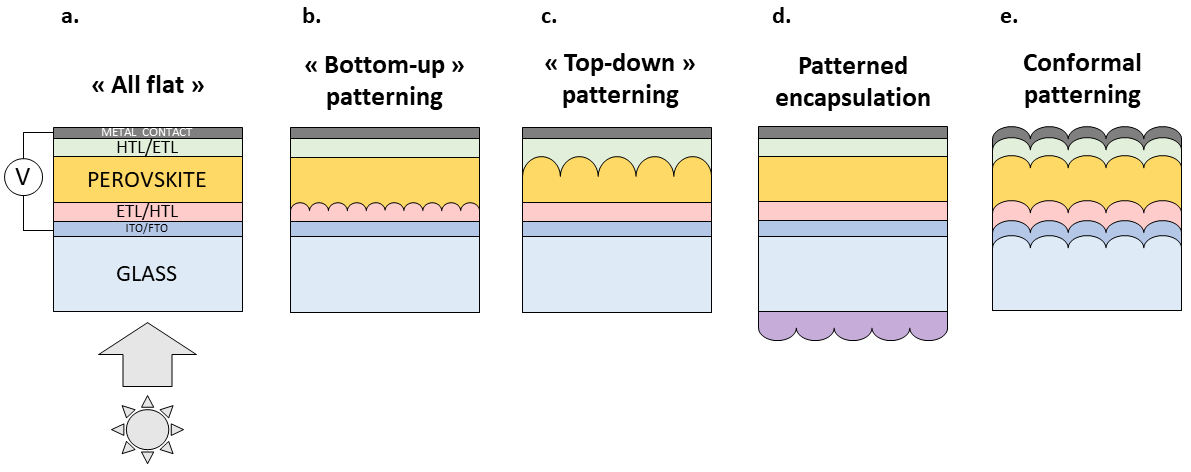}
		\caption{Various typical architectures of complete PPSC obtained using nanofabrication. a. Almost flat (possible nanoscale roughness) b. patterned perovskite using "bottom-up" approach, i.e. structuration of the ETL/HTL, then overcoating with perovskite; c. patterned perovskite using "top down" approach, i.e. perovskite coating then structuration and overcoating. d. Flat perovskite but patterned surrounding layers and e. Patterned substrate coated with conformal stack. All envisaged patterned can be either periodic or aperiodic.}
		\label{fig:Architecture}
	\end{figure}
	
	Let us present an overview of the main patterning processes and deriving architectures that have been proposed in the literature.
	Due to the low diffusion lengths in the first fabricated metal-halide hybrid perovskites, first PPSC used mesoporous architectures to limit recombination. Indeed, TiO$_2$ scaffolds were infiltrated with the perovskite material, enabling a more efficient collection of the carriers, and, in some cases, LM \cite{Lee2015,Lin2016,Ramos2016}. Nevertheless, as already mentioned in the introduction, most promising approaches relies nowadays in planar architectures. 
	
	In the following, layers will be considered as almost flat provided their roughness is at a sub-wavelength, nanometre scale, especially thanks to a good crystallinity in the case of the perovskite. 
	
	However, a 2D structuration can be introduced in the PPSC stack, notably for single junction PPSC. In this frame, the perovskite layer itself can be one of the patterned layers, or only surrounding layers are structured, and the perovskite can be considered as flat. 
	The various envisaged technical solutions within each of the two strategies are sketched in Figure \ref{fig:Architecture} b-e.
	Patterning of the perovskite layer can be achieved by various approaches. Such a structuration can first be generated by depositing this active layer on top of patterned layers, for example the ETL/HTL standing underneath ("bottom-up" like approach, Figure \ref{fig:Architecture} b.); the pattern can be either random \cite{Xu2020,Huang2016,Pascoe2016}, or periodic \cite{Paetzold2015,Abdelraouf2016,Wang2018,Kim2019}. Alternatively, the perovskite layer itself can be directly patterned \textit{after its deposition} ("top-down" like approach, Figure \ref{fig:Architecture} c.), again either in a random \cite{Wang2018a,Wang2018b} or periodic way \cite{Schmager2019,Schmager2019b}. 
	
	Another approach is to realize flat perovskite layers and to corrugate other adjacent layers (Figure \ref{fig:Architecture} d.), such as the encapsulation layer \cite{Nguyen2016,Li2019,kim_light_2021}, transport layers \cite{kang2016b,Qarony2018,wang_coordinating_2021}, possibly associated with the metal back contact \cite{Wei2017,Zhang2018b}, or only the front contacts \cite{Hossain2020,Tockhorn2020}, and finally a glass substrate covering layer \cite{Tavakoli2015,Dudem2016,Jost2017,Peer2017,doi:10.1002/adom.201900018,Thangavel2020}.
	
	Finally, the front substrate or the back contact layer can be patterned within a broad range of sizes, from the hundreds of nanometers to almost the millimeter; then the other layers of the stack are conformally deposited, leading to a fully structured cell \cite{Qarony2018,Wang2016b,Du2016,Shi2017,Soldera2018,Tooghi2020a,Wang2019,Soldera2020,Haque2020} (Figure \ref{fig:Architecture} e.). It is noticeable that among these numerous papers, only two \cite{Tockhorn2020,Soldera2020} are dealing with experimental results. Indeed, spin coating of perovskite on a patterned substrate remains challenging. The former mesoporous architecture has recently been revisited under the form of opal like structuring \cite{Lobet2020}.
	
	\subsection{Influence of the choice of Materials}
	
	\subsubsection{Perovskite medium}
	The choice of the absorbing material for PPSC results from multiple constraints. It should first exhibit an optimal band gap energy, in the range 1.1 - 1.4 $eV$\cite{shockley1961detailed}, as well as optimal electrical properties and stability. Then, various deposition processes can be envisaged for a given metal halide perovskite. It may mainly lead to various grain sizes of the material during the crystallization process. For sizes of the order of the wavelength, this can affect LM. 
	
	Methylammonium lead iodide perovskite material (CH$_3$NH$_3$PbI$_3$, MAPI) has been widely considered in the early stages of PPSC development, thanks to its rather simple chemical composition, and despite $E_g$ slightly above the optimal range. 
	It is even noticeable that different $E_g$ are reported for MAPI typically from 1.55 to 1.6 $eV$ (see e.g. \cite{jiang2015}). This implies a more than 7\% uncertainty on the maximum achievable $J_{sc}$, that is already of the order of the achievable gain using LM as reported in the following. Thus, the quantitative comparison between different studies, even implying the same material within the meaning of the chemical composition, is rather delicate. Differences of aging of the perovskite could also impact the comparison. Finally, various dispersion models can be used to fit the dielectric function \cite{Li2020b}, leading to differences that also impact the optical indices that are used for simulations. 
	
	For all these reasons, it has to be emphasized that, in the following, only the relative impact of LM could be put into perspective, but the various given performances cannot be accurately compared. Finally, if MAPI can still be used as a typical case study, other materials have been developed in the past few years. They are more stable and exhibit a slightly lower $E_g$, mainly thanks to the substitution of MA by FA (Formamidinium, (NH$_2$)$_2$CH), possibly with Cs, as can be seen later.
	
	\subsubsection{Transport and contact layer materials}
	For the transport layers, mainly polymer materials are used, as well as thin layers of inorganic materials such as TiO$_2$. It is noticeable than materials such as those based on C$_{60}$ fullerene, acting as ETL, exhibit significant absorption at the shortest wavelengths of the solar spectrum, whereas polymers used as HTL mainly absorb at larger wavelengths, close to and above the band gap of the perovskite. Then, there is often a compromise to be found between suitable band levels, high electrical conductivity, and barrier to possible migrations, and low parasitic absorption, even if the thicknesses of these transport layers remain limited.
	
	\section{Photonic engineering}
	\label{Photonic concepts description}
	
	\begin{figure}[b]
		\begin{center}
			\includegraphics[width=\linewidth]{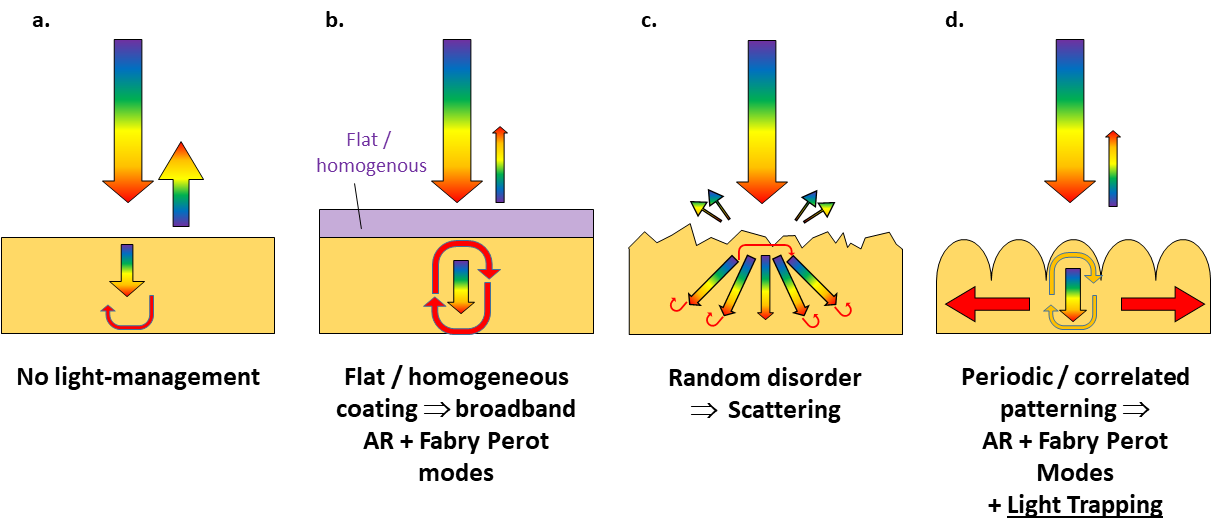}
			\caption{Main Light Management strategies a. Without peculiar LM, reflectance can be important. b. Using a flat structure, or a patterned that can be considered as homogeneous (e.g. given its subwavelength dimensions), LM can lead to a rather broadband AR effect as well as to Fabry Perot modes that can enhance the absorption. c. Using a random texturation, LM can results from scattering, possibly Lambertian, that can also enhance the absorption. d. Using a periodic or strongly correlated patterned, in addition to AR and Fabry Perot effects, some Light Trapping can occur, drastically enhancing the absorption especially at band edge of perovskite.}
			\label{fig:LightManagement}
		\end{center}
	\end{figure}
	
	Let us now summarize the various kinds of PE, mostly at the wavelength scale, that can be envisaged to improve LM in both single-junction and 2T multi-junction PPSC, and their impact on geometrical requirements in the PPSC. They are sketched in Figure \ref{fig:LightManagement}. Most of the strategies derive from those already developed for other kinds of direct or even indirect band gap material used in thin film \cite{massiot2020progress}.
	
	In this frame, LM results from various light-structure interaction mechanisms. These mechanisms occur mainly either in the vertical direction (i.e. orthogonal to the stack) or in the directions of the patterns. Depending on the scale of the structure with respect to the wavelength, three main models can be used: first, at sub-wavelength scales, an effective medium; then, at wavelength scale, wave optics, or, at larger scales, geometrical optics.
	
	\subsection{Photonic engineering in multilayer unpatterned stack}
	\label{photonic concepts multilayer}
	It is well known that geometrical optics are mainly suitable when the dimensions are far larger than the wavelength. Thus, it cannot accurately describe the interference effects that occur within the thickness of the numerous thin films that compose the cell. In contrast, the optical indices of medium textured at subwavelength scales can be homogenized in a approximated way using Effective Medium Theories such as the Bruggeman model. Then, descriptions involving wave optics are the most rigorous, and wave optics mechanisms might also be the most promising ones.
	
	Let us first briefly recall the two basic mechanisms able to enhance light harvesting in a flat stack, i.e. without the assistance of any patterning. In a rather high index, highly absorbing material, a properly designed stack enables to drastically reduce the impedance mismatch with the surrounding environment, whereas in a finite thickness absorbing layer, especially with a low absorption, it is rather a Fabry Perot like approach that can be used, leading to highly enhanced but narrow band absorption. 
	
	Within a typical PPSC stack, having a metallic contact that can act has a back mirror, these two approaches can be invoked. Schematically, at short wavelength, the large extinction coefficient of the perovskite enables the single pass absorption in the perovskite film, but AR effect is required. At larger wavelengths close to band gap, where the absorption drops, multiple passes are required, the second approach is well suited (Figure \ref{fig:LightManagement} b.). 
	
	However, the exact nature of the layers, and thus their optical indices and thicknesses are restricted by other constraints such as energy band levels, electrical resistance, risk of shunting, or even diffusion barriers for some species. Consequently, parasitic optical absorption has to be carefully studied. Moreover, broadband enhancement of the absorption under normal incidence is unlikely to be obtained using such simple architectures. 
	
	\subsection{Photonic engineering in multilayer, patterned stack}
	\label{photonic_concepts_multilayer_patterned}
	In addition to the continuum of propagative waves, light can be confined into the discrete set of in-plane, transverse guided modes existing in the stack. It is noticeable that the in-plane wave vector of any of these guided modes is larger than those of the free space modes. Thus these modes cannot be simply coupled from the free space in a perfectly flat stack, without any periodic or aperiodic corrugation. Then, the various kinds of in-plane structurations - having dimensions from the wavelength scale up to two orders larger - all induce diffraction. Whereas the impinging light lies around the normal incidence and has thus negligible in-plane wave vector $k_{in\parallel}$, the light inside the device is diffracted. Indeed, the electromagnetic field can be expanded over a set of plane waves thanks to the in-plane spatial frequencies induced in the medium by the structuration. The reflected waves also experience diffraction accordingly. Considering the various kinds of structurations and thus of the resulting spatial frequencies, three main cases can be envisaged:
	
	\begin{enumerate}[label=\roman*)]
		\item A random structuration tends to induce isotropic diffraction, so Lambertian scattering. Moreover, this scattering does not depend on the wavelength within the absorbed spectral range. It can typically lead to broadband AR effect (Figure \ref{fig:LightManagement} c.). However the absorption enhancement remains limited. Indeed, given the already large absorption (except at the band edge) of the perovskite, and the back reflection induced by the metallic contact, it will be at most of the order of $4n^2$ with an index lower than other materials, in particular inorganic ones \cite{Yablonovitch1982}.
		\item A correlated disordered structuration that enhances a specific set of spatial frequencies, leads to a spectral dependent absorption enhancement. A careful choice of the sizes of the patterns is required to enhance the absorption in the desired spectral range \cite{Vynck2012nmat}.
		\item As a particular case from the previous case, periodic, so-called photonic crystal (PC) structurations can be envisaged. Most of them consists in a square or triangular lattice a simple pattern such as pillars or holes with a vertical profile, or smoother pyramidal, or even parabolic profiles. They mainly lead to discrete modal properties. A few complex patterns arranged in a periodic way have been explored \cite{martins2013deterministic,Oskooi2014,Bozzola2014,ding2016design,Li2020}, enabling an absorption enhancement in a targeted spectral range thanks to a larger local density of modes. 
	\end{enumerate}

	In any case, the diffraction efficiency, i.e. the amount of light that is effectively diffracted, and that is thus intended to be absorbed, strongly depends on the pattern of the structuration, including the optical index contrast, as well as on the diffraction order $p$. Whereas, as already mentioned, for scattering, the diffraction efficiency hardly depends on the direction, for diffraction by periodic structures, it is generally larger for the first diffraction order.
	
	In this frame, it is noticeable that a LT phenomenon rigorously occurs when the impinging light is coupled thanks to, at least, strongly correlated structure, ideally periodic patterns, having respectively a correlation length or a period $\Lambda$, within at least one guided mode of the stack (Figure \ref{fig:LightManagement} d.). According to a perturbation approach \cite{CamarilloAbad2020}, in a 1D case for sake of illustration without losing generality, a phase matching condition can be written under the form:
	
	\begin{equation}
		\beta_m=k_{in\parallel}+p \frac{2\pi}{\Lambda}
		\label{QGM}
	\end{equation}
	Where $\beta_m$ is the in-plane wave number of the $m^{th}$ guided mode of the stack. Due to the time reversal symmetry, the light coupled into the guided mode, if not fully absorbed, can be decoupled after a certain length. These modes are thus so-called "quasi-guided mode" (QGM).
	
	Such modes can drastically enhance the light path inside the waveguide layer. Thus, if the in-coupled QGM is mainly confined in the perovskite layer rather than in the surrounding layers, the useful absorption is enhanced. Moreover, it also depends on the absorption; more precisely, the absorption is optimized at critical coupling, i.e. when an equilibrium is reached between diffraction efficiency and absorption. Consequently, the spectral bandwidth of the LT is set since the quality factor of the mode at the optimal absorption is of the order of $n/2\kappa$, $n,\kappa$ being respectively the real and imaginary parts of the complex optical index \cite{Park2009}.
	
	Such a resonant LT appears promising to enhance the low absorption close to the band gap of the perovskite, provided it is not reduced elsewhere at shorter wavelengths. More precisely, within the width of the resonance, the absorption can be significantly larger than the previously mentioned broadband limit \cite{Yu2010}. It could also be used in tandem PPSC mainly to optimize the $J_{sc}$, especially provided LT occurs in a guided mode specifically confined in one of the perovskite layers. As already discussed, for the other spatial frequencies that do not lead to LT, patterning can still result in a rather broadband AR effect, in addition to the one resulting from the design of the stack. Further analysis of all these effects can be found elsewhere \cite{Zanotto2010,Brongersma2014}. 
	
	In addition to these PE approaches that contribute at first order to enhance the absorption and so the $J_{sc}$, other strategies could be used to enhance the $V_{oc}$, using a PR mechanism. In a simplified way, PR is supposed to induce at the $V_{oc}$ operating point a high density of photons into the absorbing materials, and so to give a chance to electrons-holes pairs to recombine radiatively rather than in a non radiative way. Typically, this requires at first order to inhibit the radiation, and consequently the absorption, in the luminescence spectral domain of the semiconductor. Thus, enhancing simultaneously the $J_{sc}$ using LT and the $V_{oc}$ using PR is a compromise \cite{Sandhu2013}.
	
	In any case, it has to be emphasized that all these effects should ideally target the perovskite layer only and not enhance the parasitic absorption in the other layers.
	
	\subsection{Main geometrical requirements on patterns for Light Trapping}
	
	As mentioned previously, 2D in-plane structurations are able to diffract the impinging sunlight under a quasi-normal incidence into the large number of modes of the PPSC, possibly including the guided modes.
	Indeed, the architectures of both single and multi-junction PPSC reviewed in the section \ref{possible_architecture} can be schematized as one or several sub-micron thick high index perovskite layers lying between two lower indices ETL and HTL and possibly also lower index layers in the multi-junction case. The whole is lying on medium index TCO on a low index glass substrate and coated on top with metal.
	Such structures exhibit several guided modes in both TE and TM polarization states.
	Let us first make the assumption of weak corrugation that doesn't change significantly the effective indices of the stack of active layers lying on a substrate. The largest effective index modes - typically 2.2 - are mainly confined in the perovskite, whereas the other modes, whose effective indices are below 1.9, can be confined in the whole stack of layers.
	In this frame, for normal incident light ($k_{in//}=0$) and for the highest efficient coupling at order $p=1$, equation (\ref{QGM}) can be rewritten under the form:
	\begin{equation}
		\beta_m=\frac{2\pi N_{eff,m}}{\lambda}=\frac{2\pi}{\Lambda}
		\label{phase_matching_equation}
	\end{equation}
	where $\Lambda$ is the period, and could even be a characteristic length in correlated patterns. This 1D case can easily be extended to 2D patterns. 
	
	Thus, to be able to efficiently couple into the fundamental guided mode, which is mainly confined into the perovskite, at wavelengths corresponding to its band edge, e.g. 750 $nm$, $\Lambda$ should be around 340 $nm$, whereas it should be larger than 400 $nm$ for higher order guided modes. This results in a pseudo-guided mode, as discussed at the end of section \ref{photonic_concepts_multilayer_patterned}. At shorter wavelengths, such structurations can also couple light into the various low effective index modes (including Fabry Perot like modes) of the stack, that may also result in a broadband AR effect. 
	This typically implies that all the micron scale or even larger patterns in any of the layers of the stack are rather unlikely to couple the impinging light into guided modes in the low absorption domain, where such a coupling is of main interest. Such patterns only induce coupling into the Fabry Perot like modes of the stack and/or diffraction at high orders with a limited efficiency.
	
	\section{Examples of reported light management strategies in PPSC}
	
	We report here on some studies to illustrate concrete LM strategies. For any case, the chosen perovskite material and its thickness are of primary importance. First of all, planar PPSC as envisaged in the section \ref{possible_architecture} for their architecture and in section \ref{photonic concepts multilayer} for the corresponding PE are discussed, putting mainly into evidence the importance of the thicknesses on LM. Then, results on in-plane patterned PPSC are reviewed, with focuses on the involved PE.
	
	\subsection{Optimization of the thicknesses in planar PPSC}
	
	Most studies related to thickness optimization only focus on that of the perovskite layer, and report mainly on the derived useful absorption, as detailed just below. However, the whole stack has been considered in a few papers, with a view to investigate the parasitic absorption, as discussed later.
	
	\subsubsection{Choice of the perovskite thickness in single junction PPSC}
	
\begin{figure}[h]
		\centering
		\includegraphics[width=\linewidth]{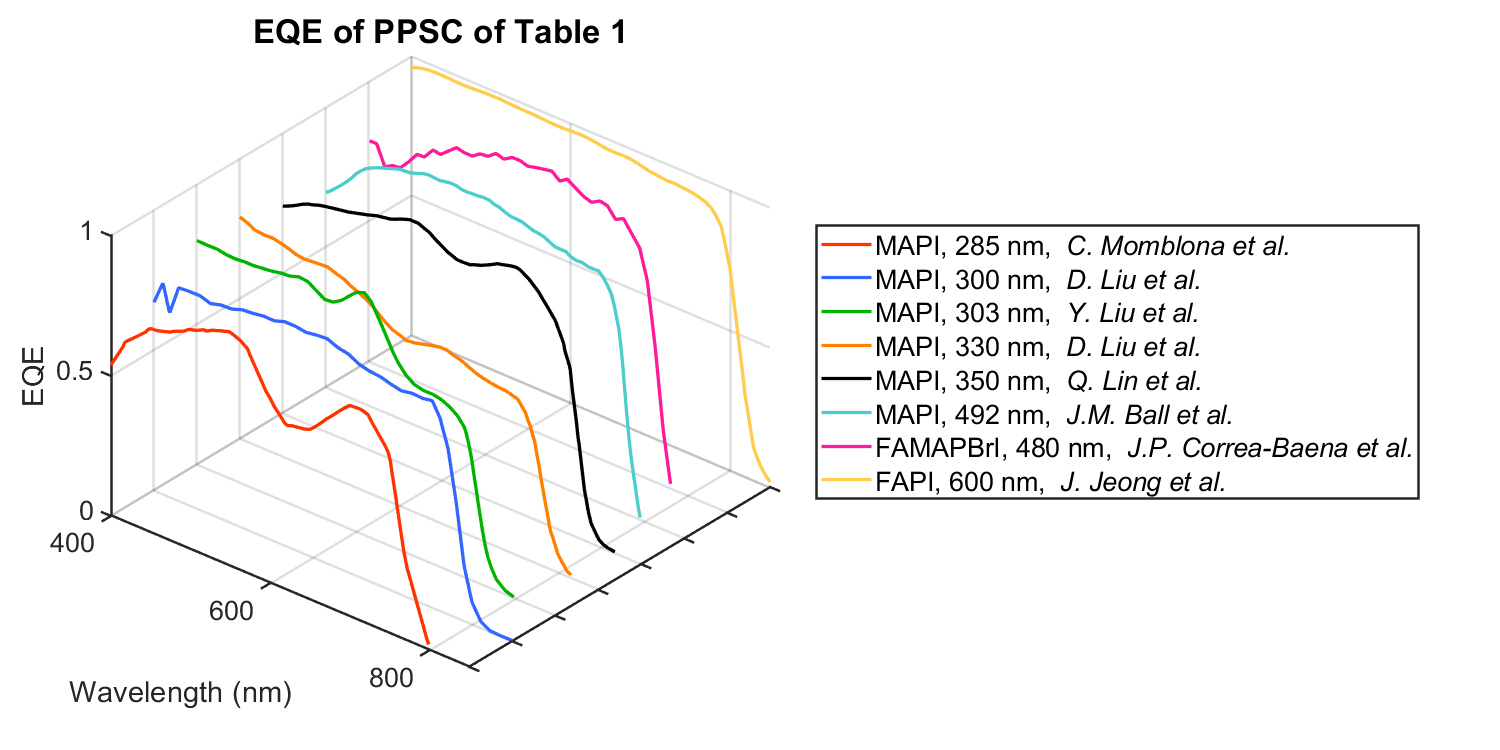}
		\caption{EQE of various PPSC also reported in Table \ref{tableau 1} \cite{Liu2014b,Momblona2014,Liu2014,Liu2019b,Correa-Baena2016,jeong_pseudo-halide_2021,Ball2015,Lin2015}.}
		\label{fig:EQE_tableau_1}
	\end{figure}

	As already mentioned, the optimal thickness of the sole perovskite layer is chosen as a compromise between maximizing optical absorption and keeping low bulk recombination of the carriers. In addition to the analysis of a few seminal cases of single junction PPSC, we present a synthesis of several studies, mainly experimental, that report on the effect of the perovskite layer thickness of planar single junction PPSC, first and mainly made of MAPI, then of other lead-halide perovskites. The detailed performances of the representative cells are summarized in Table \ref{tableau 1} and, when available, the EQE are plotted in Figure \ref{fig:EQE_tableau_1} in a spectral range limited from 400 to 800 $nm$ for sake of simple comparison.\\
	
	\paragraph{MAPI}
	
	One of the first high efficiency planar PPSC \cite{Liu2013} which is made of an about 300 $nm$ thick MAPICl layer exhibited a $J_{sc}$ of 21.5 $mA.cm^{-2}$. In this seminal paper, M. Liu \textit{et al.} already mentioned the possibility of an optimum thickness due to the balance between absorption and recombination. It was shown later by Y. Da \textit{et al.} \cite{Da2018} that this cell suffered from optical losses that represented 11.3\% of the energy losses, and especially more than 20\% of the achievable $J_{sc}$. This was mainly related to reflectance and parasitic absorption in the FTO. This thus emphasizes the importance of minimizing these effects by a careful design of the stack.
	
	Then, the first high-efficiency PPSC, solution processed at room temperature, proposed by D. Liu \textit{et al.} \cite{Liu2014b} was made of an about 300 $nm$ thick MAPI layer. A $J_{sc}$ of 20.4 $mA.cm^{-2}$ was obtained, so about 5 $mA.cm^{-2}$ below the maximum achievable $J_{sc}$ given a typical $E_g$ of MAPI. The EQE (see Figure \ref{fig:EQE_tableau_1}) was rather flat and remained lower than 0.8. The deficit could be due to both optical losses (too low absorption in the perovskite) and electrical losses (too high recombination). 
	
	C. Momblona \textit{et al.} \cite{Momblona2014} conducted an experimental study on PPSC made of MAPI deposited by thermal evaporation, especially by increasing its thickness from 160 to 900 $nm$. If the $J_{sc}$ increased rather monotonously with the perovskite layer thickness, it never exceeded 20.4 $mA.cm^{-2}$. The main parameter that decreased with increasing perovskite layer thickness was the fill factor. As a result, the cell owning optimal performance was obtained for a perovskite layer thickness around 300 $nm$. More precisely, whereas the $J_{sc}$ slightly increases, both the $V_{oc}$ and the $FF$ significantly decreased at thicknesses larger than 300 $nm$. At these early ages of PPSC, the corresponding optimal PCE was 12.7\%, with a $J_{sc}$ of 18.8 $mA.cm^{-2}$. Then, leading further investigations on the $J_{sc}$ deficit, it could be derived from the EQE (see Figure \ref{fig:EQE_tableau_1}) that optical losses in almost 300 $nm$ thick perovskite were important at both short wavelengths and from 600 to 700 $nm$, compared to the 900 $nm$ thick perovskite layer that fully absorbs. The enhanced absorption in the range 700 to 750 $nm$ is likely due to a cavity effect. Thus, even if such a thickness of about 300 $nm$ appears to be optimal under the PCE point of view, it remains too thin to avoid optical losses. In addition, even with such thin absorbing layers, such cells still suffers from electrical losses. 
	
	In a similar way, D. Liu \textit{et al.} \cite{Liu2014} also then studied PPSC consisting in a ITO/ZnO/MAPI/P3HT/Ag stack, in which MAPI was obtained using thermal evaporation or solution processing. Again, in spite of a larger absorption and limited short-circuit for thicker films, a limited diffusion length led to an optimum thickness of the MAPI layer of about 330 $nm$ for the solution processed PPSC. It is noticeable that the corresponding EQE (see Figure \ref{fig:EQE_tableau_1}) was rather low and flat. This could be related to scattering due to sub-micron crystallinity that mitigates cavity effects at larger wavelengths.
	
	More recently, Y. Liu \textit{et al.} \cite{Liu2019b}, still using MAPI, similarly concluded that there were more power loss in PPSC when the thickness of the MAPI layer was either less or more than about 300 $nm$. They noticed that the grain size increased with the thickness, in favor of a larger thickness. However, it was noticeable that, at wavelengths larger than 650 $nm$, the EQE of the 303 $nm$ thick MAPI PPSC (see Figure \ref{fig:EQE_tableau_1}) was lower than the one of the 564 $nm$ thick MAPI PPSC. This is related to too low absorption in this spectral range. \\

	\paragraph{Other lead-halide perovskites}
	
	As an alternative, J.P. Correa-Baena \textit{et al.} \cite{Correa-Baena2016} used multi-cations perovskites materials; these were recently under the spotlight, especially considering their better stability \cite{jeon2015compositional}. Using FA$_{0.83}$MA$_{0.17}$Pb(I$_{0.83}$Br$_{0.17}$)$_3$ together with mesoporous TiO$_2$, the thickness that maximises the PCE was found to be at least 480 $nm$. The corresponding $J_{sc}$ was almost 24.4 $mA.cm^{-2}$. Considering the band gap of this material, reported to be about 1.63 eV \cite{jacobsson2016exploration}, this is already the maximum achievable value. The EQE (see Figure \ref{fig:EQE_tableau_1}) is noticeably high (also thanks to the an IQE of almost 100\%) and flat at large wavelength. This could be related to the mesoporous TiO$_2$ that induces some scattering effect. 
	
	Unlike the previous papers, M. Rai \textit{et al.} \cite{Rai2020} focused on the $V_{oc}$ deficit related to non-radiative recombination as a function of the thickness, expressed in terms of molar concentration of the precursor solution. This time, the studied perovskite was Cs$_{0.2}$FA$_{0.8}$Pb(I$_{0.85}$Br$_{0.15}$)$_3$, whose band gap has been found to be 1.62 $eV$, so close to the one of MAPI. They noticed as well that the $V_{oc}$ decreases with the thickness, whereas the $J_{sc}$ increases. As for the previously mentioned paper, the grain size increases with the thickness; the molar concentration of the precursor solution that maximizes PCE correspond to thickness of the perovskite layer of about 400 $nm$, so 30\% larger than the one usually obtained using MAPI. This might be due to a larger diffusion length of the photocarriers for this perovskite.
	
	K.B. Nine \textit{et al.} \cite{nine_performance_2020} simulated optically and electrically the effect of the thickness of several FACsPI layers, leading to a larger optimal thickness, of almost 600 $nm$, so twice the thickness usually found using MAPI. Indeed, better electrical properties of this perovskite material, especially carrier mobility, allow such a thicker layer. \\
	
	\paragraph{Synthesis}

	\begin{table}[!h]
		\begin{tabular}{@{}lllllll@{}}
			\hline
			Material & Thickness ($nm$) & $J_{sc} (mA.cm^{-2})$ & $V_{oc} (V)$ & $FF$ & PCE (\%) &  Ref \\
			\hline
			MAPICl & 330 & 21.5 & 1.07 & 0.67 & 15.4 & \cite{Liu2013} \\  
			MAPI & 300  & 20.4 & 1.03 & 0.749 & 15.7 & \cite{Liu2014b} \\    
			MAPI & 285  & 18.8 & 1.07 & 0.63 & 12.7 & \cite{Momblona2014} \\
			MAPI (solution process) & 330  & 17 & 0.94 & 0.62 & 11.8 & \cite{Liu2014} \\
			MAPI & 303 & 21.3 & 1.07 & 0.715 & 18.4 & \cite{Liu2019b} \\
			FAMAPBrI & 480 & 24 & 1.14 & 0.75 & 20.8 & \cite{Correa-Baena2016} \\
			FAPI & 600 & 26.35 & 1.189 & 0.817 & 25.59 & \cite{jeong_pseudo-halide_2021}\\
			MAPI & 492 & 21.56 & nr & nr & nr & \cite{Ball2015}\\
			MAPI & 350 & 21.9 & 1.05 & 0.72 & 16.5 & \cite{Lin2015} \\  
			\hline
		\end{tabular}
		\caption{Experimental performances of planar PPSC with optimized perovskite thickness (nr: not reported)}
		\label{tableau 1}
	\end{table}
	
	It is noticeable that these studies tend to an optimal thickness of about 300 $nm$ when using MAPI. This thickness could lead to state-of-the-art PCE of J. Li \textit{et al.} \cite{LI2018331}, with noticeably high $J_{sc}$ of 24.1 $mA.cm^{-2}$, mainly thanks to an ETL including graphdiyne, which improves electrical properties. Then, FAPI is the perovskite used for the best up-to-date single junction PPSC \cite{jeong_pseudo-halide_2021} which performance were mentioned in the introduction, with a thickness of about 600 $nm$, even if, among numerous refinements of the architecture, the rough, about 50 $nm$ thick TiO$_2$ layer already may help to trap the light. 
	
	It appears that choosing a thickness smaller than the one maximizing the $J_{sc}$ could help to keep low resistance, so high FF and PCE. Then, with such a $J_{sc}$ around 90\% of the maximum achievable $J_{sc}$, LM can optimize the absorption in the perovskite, especially at photon energies close to the band gap, where the absorption starts to decrease.
	
	In any case, it is of high interest to understand the reasons why large variations of the internal perovskite absorption spectrum have been obtained for similar thicknesses of the same perovskite material (MAPI), as can be noticed through the cases 2 to 5 of Table \ref{tableau 1} and the corresponding EQE plotted in Figure \ref{fig:EQE_tableau_1}. They might be due to differences in the surroundings layers (thicknesses, indices) or in the perovskite itself, like microcrystallinity that can induce scattering. These phenomena will be described in the following, starting with the analysis of the effect of the surroundings layers in a unpatterned stack. 
	
	\subsubsection{Single junction unpatterned PPSC stack analysis}
	If there can be an optimal thickness for the perovskite layer, the effect of the other layers, especially on the absorption and reflectance, should also be analyzed. Here, we review a selected set of publications focusing on this issue.
	
	J.M. Ball \textit{et al.} \cite{Ball2015} studied planar PPSC using an optical model based on the transfer-matrix formalism with experimentally determined complex refractive index data. They focused on a typical stack made of FTO/TiO$_2$/MAPI/SPIRO-OMeTAD/Au. Under the assumption of $E_g = 1.56 eV$, the detailed analysis revealed that for a calculated $J_{sc}$ of 21.56 $mA.cm^{-2}$, parasitic absorption induces a lack of about 1.72 $mA.cm^{-2}$, and reflection losses corresponding to a $J_{sc}$ decrease of 1.36 $mA.cm^{-2}$, so a total of more than 10\% of the collected current. On the other hand, the IQE losses can be estimated to another 10\%. Moreover, most of the optical losses take place in the 420 $nm$ thick FTO. This underlines the importance of taking care of all the layers of the stack.
	
	\begin{figure}[h]
		\centering
		\begin{subfigure}[b]{0.45\textwidth}
			\centering
			\includegraphics[width=\textwidth]{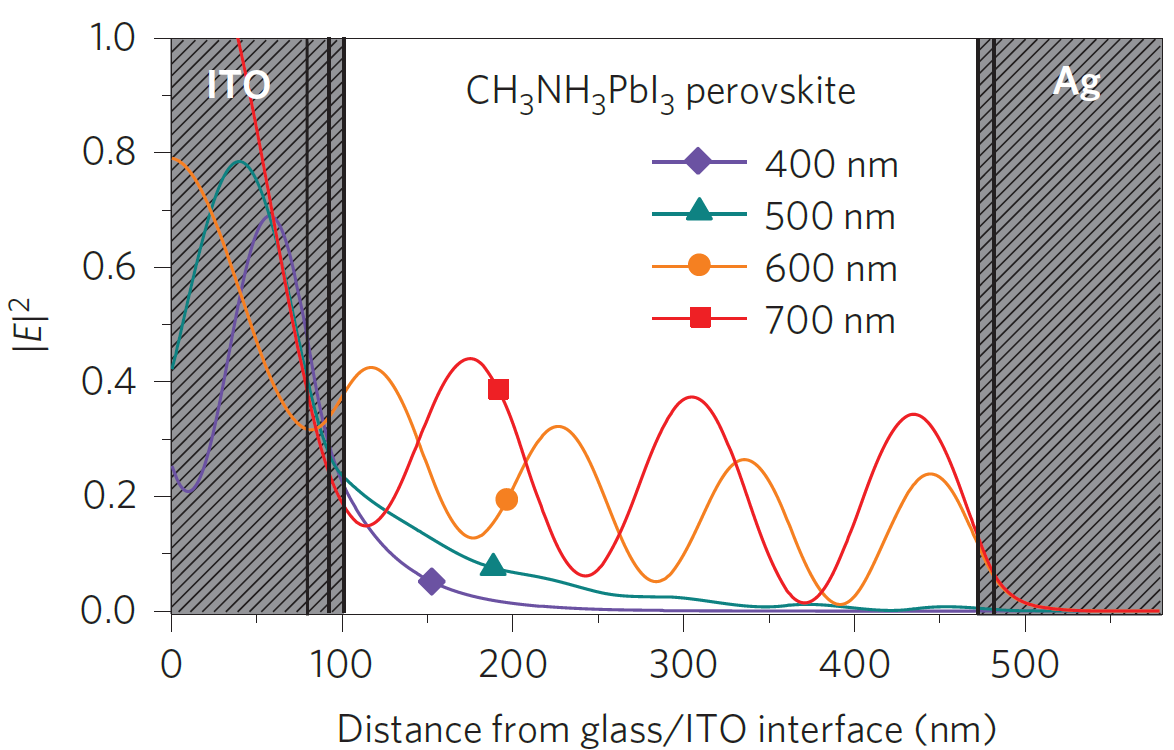}
			\caption{Optical-field distribution for four wavelengths: for $\lambda < 500 nm$ the optical-field distribution follows the Beer–Lambert law and no optical field reaches to the back electrode as a result of the high absorption coefficient. In such cases the absorption is saturated and no optical interference occurs. For $\lambda > 500 nm$ the optical field is governed by low-finesse cavity interference.}
		\end{subfigure}
		\hfill
		\begin{subfigure}[b]{0.45\textwidth}
			\centering
			\includegraphics[width=\textwidth]
			{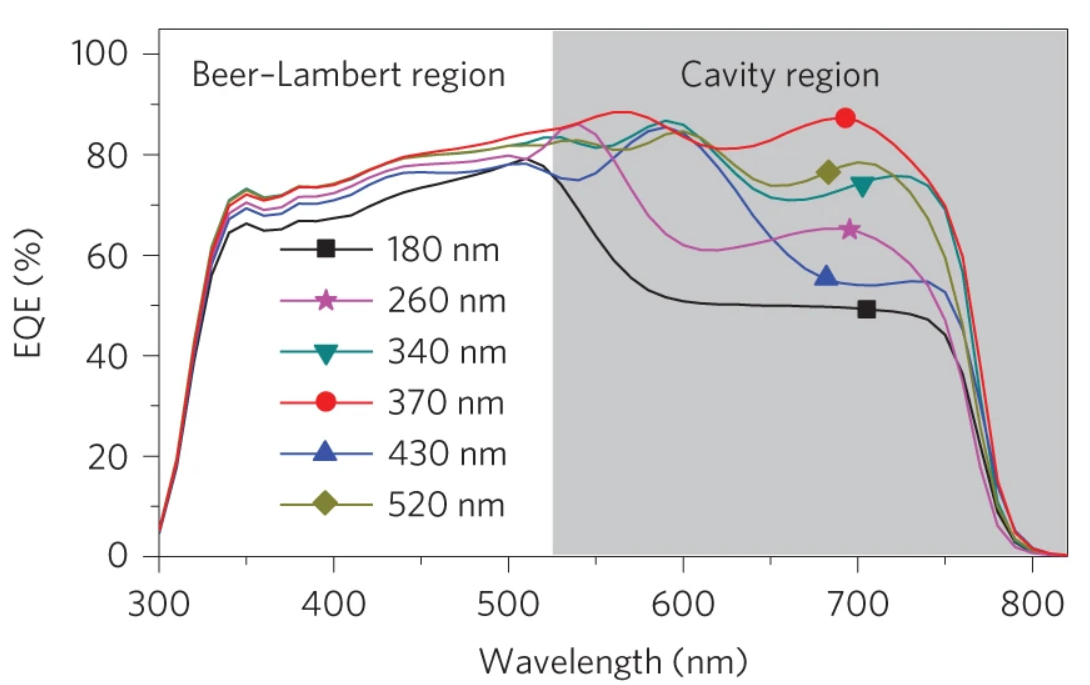}
			\caption{EQE spectra of devices with different MAPI-layer thicknesses. For $\lambda < 500 nm$ there is minimal influence of the film thickness, but for $\lambda < 500 nm$ the EQE is strongly thickness dependent because of the optical interference (Fabry Perot like modes)}
		\end{subfigure}
		\caption{PPSC properties as a function of the MAPI layer thickness \cite{Lin2015}. Reproduced with permission}
		\label{fig:lin eqe}
	\end{figure}
	
	Q. Lin \textit{et al.} \cite{Lin2015} deeply investigated the properties of PPSC made of MAPI. Thanks to their measurement of the optical indices of the MAPI and of the other materials, they simulated the properties of PPSC, assuming an IQE of 100\%. They identified the two absorption regimes of a flat PPSC, previously discussed in section \ref{photonic concepts multilayer}, namely the single pass and the cavity regime. Thus, thanks to optical cavity effects at wavelengths larger than 500 $nm$, the noticeably high EQE of their cell (see Figure \ref{fig:EQE_tableau_1}) appeared to be optimized for a thickness of about 370 $nm$ (see Figure \ref{fig:lin eqe}). The experimental data noticeably confirmed the simulation study, since a cell having a 350 $nm$ thick MAPI layer led to the best properties, $J_{sc}=21.9 mA.cm^{-2}$, $V_{oc}=1.05 V$, $FF=0.72$ and $PCE=16.5\%$. Given the resulting limited thickness of the MAPI, the obtained $J_{sc}$ was 87\% of the maximum achievable value. These remarkable results were also obtained thanks to the optimization of the p and n type interlayers that were used to optimize the heterojunctions. It was especially shown that the PCDTBT p type interlayer has to be as thin as possible ( $\sim 5 nm$), because of its absorption, even if it enabled a nice crystallinity of the MAPI. Otherwise, given the almost 100\% IQE, most of the 4 to 5 $mA.cm^{-2}$ decrease of the photocurrent (depending on the exact value of the band gap energy of the considered MAPI) corresponds to optical losses, which could be either parasitic absorption or reflectance.
	
	M. Van Eerden \textit{et al.} \cite{VanEerden2017} analyzed the optical losses of 350 $nm$ thick MAPI layers PPSC. Within the considered cell, most of the losses are related to reflectance (equivalent to 4.2 $mA.cm^{-2}$), whereas the parasitic absorption remains limited to the half, and despite a subwavelength roughness (typically from 10 to 50 nm RMS) at both ITO/TiO$_2$ and MAPI/SPIRO-OMeTAD interfaces. 
	
	\subsubsection{Analysis of unpatterned, up-to-date record 2T tandem PPSC}
	\label{tandem_plan}
	
	\begin{figure}[h]
		\centering
		\begin{subfigure}[b]{0.3\textwidth}
			\centering
			\includegraphics[width=\textwidth]{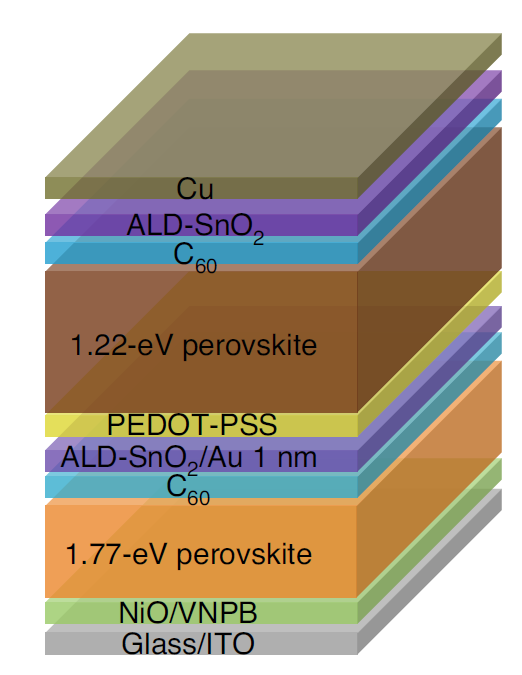}
			\caption{Stack of materials }
			\label{fig:xiao tandem schema}
		\end{subfigure}
		\hfill
		\begin{subfigure}[b]{0.3\textwidth}
			\centering
			\includegraphics[width=\textwidth]{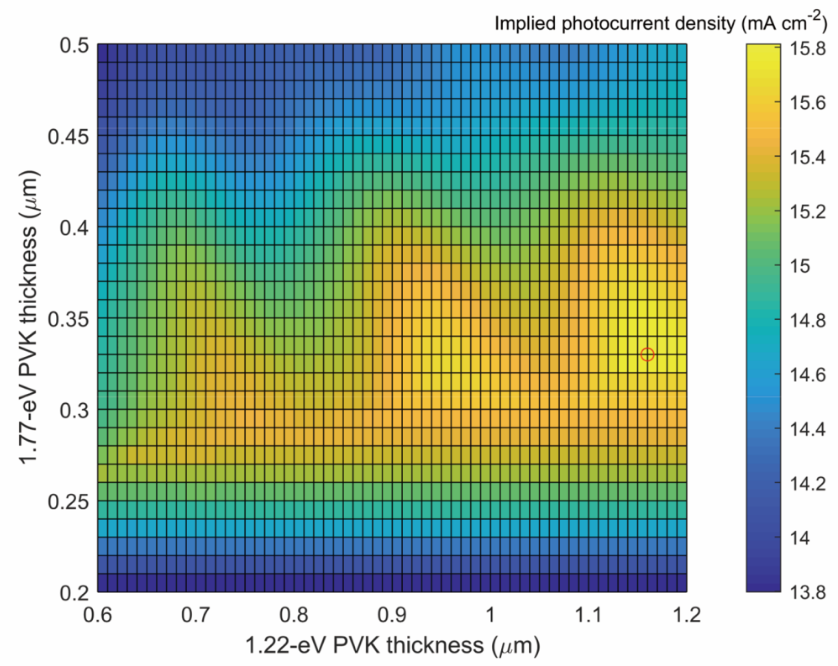}
			\caption{Simulated $J_{sc}$ as function of wide-band gap and narrow-band gap perovskite layer thicknesses.}
			\label{fig:xiao tandem jsc }
		\end{subfigure}
		\hfill
		\begin{subfigure}[b]{0.3\textwidth}
			\centering
			\includegraphics[width=\textwidth]{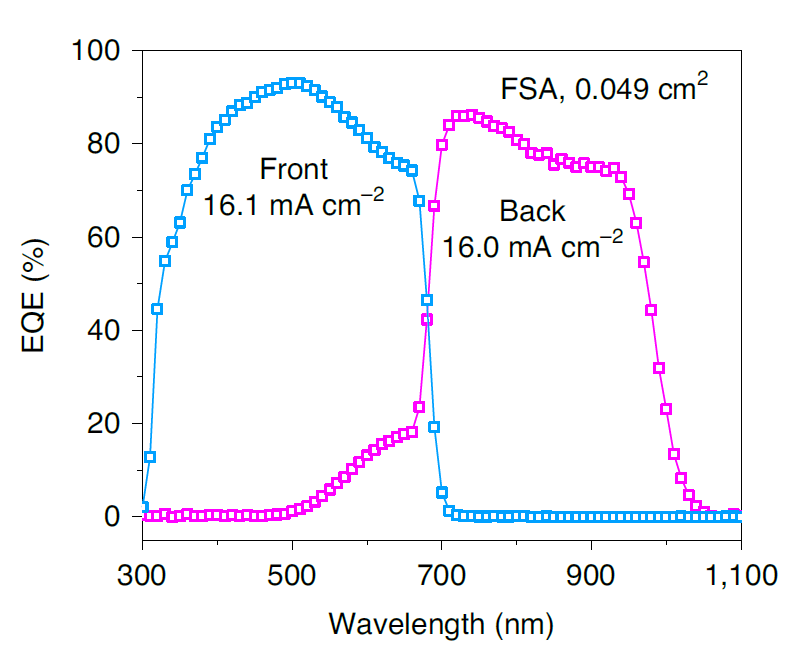}
			\caption{EQE of both subcells.}
			\label{fig:xiao tandem eqe }
		\end{subfigure}
		\caption{Best up-to-date 2T tandem, all perovskite PPSC \cite{xiao_all-perovskite_2020}. Reproduced with permission.}    
	\end{figure}
	
	When compared to other tandem architectures, the 2T case leads to possibly limited optical losses, but such cells require an equilibrium of the current densities delivered by the two subcells. Among recent results \cite{wang_prospects_2021}, K. Xiao \textit{et al.} \cite{xiao_all-perovskite_2020} have deeply investigated 2T tandem PPSC (see Figure \ref{fig:xiao tandem schema}) both numerically and experimentally. They have shown using optical simulations that the reachable $J_{sc}$, which is indeed the smallest $J_{sc}$ delivered by each of the subcell, strongly depends on the thicknesses of each perovskite layer (see Figure \ref{fig:xiao tandem jsc }), provided that thicknesses of the other layers have been set taking into account both electrical and optical properties (low parasitic absorption). Indeed, it appears that the top perovskite has to be thick enough to absorb enough light, but that a too thick layer also absorbs too much light, and finally that the overall $J_{sc}$ is limited by the bottom sub cell. The thickness of the perovskite bottom sub cell has to be large enough, more than 1.1 $\mu m$.
	When fabricated on a small surface, such a cell exhibits even a slightly larger $J_{sc}$ than simulated, likely due to the too pessimistic predicted absorption. Even if this result remains a record at the time of its publication, its EQE (see Figure \ref{fig:xiao tandem eqe }) reveals some non-idealities. Indeed, in addition to the relatively low values of the EQE plateaus, there is an overlap of the absorption domains of the two perovskite materials between 500 and 700 $nm$. This means that some undesirable thermalization still occurs below 700 $nm$ in the bottom sub cell. 
	
	\subsection{In-plane structuration for light management in single junction PPSC}
	\label{light_management}
	
	At this stage, it appears that even the best reported cells exhibit an EQE that can still be improved using PE at the wavelength scale, even if most of them already benefit from the roughness of the microcrystalline perovskite.
	
	In this section, we will highlight important criteria that should be met to ensure a fair demonstration of LM. The report on selected studies is organized as a function of the various architectures as envisaged in section \ref{possible_architecture}.
	
	\subsubsection{General criteria for fair LM demonstration}
	The impact of LM on the performance of a solar cell can be demonstrated by comparing a patterned device with an unpatterned reference. In case of superficial structurations or with the addition of specific LM layers, the absorption improvement appears generally obvious compared to the same unpatterned stack, or, even more favorable, to the unpatterned stack without the additional flat LM layer. Moreover, integrating LM layer could lead to a better charge collection, by decreasing the carrier path; the specific LM effect would then be difficult to distinguish. Therefore, great care should be taken in the definition of a reference structure or device. In particular, the volume of perovskite materials should be as close as possible in both structures. Additionally, the net enhancement of EQE, efficiency or even yield value can be accurately estimated only if the unpatterned reference has been optimized first. These conditions should be met in order to discuss LM for PPSC performance optimization. 
	
	\subsubsection{Periodic patterning for resonant LT}
	\paragraph{Superficial structuration and light management layers}
	
	\begin{figure}[h]
		\centering
		\begin{subfigure}[b]{0.45\textwidth}
			\centering
			\includegraphics[width=\textwidth]{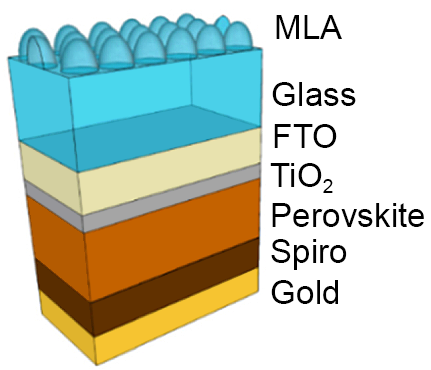}
			\caption{Schematic of the PPSC.}
			\label{fig:peer bismwas_schema}
		\end{subfigure}
		\hfill
		\begin{subfigure}[b]{0.45\textwidth}
			\centering
			\includegraphics[width=\textwidth]{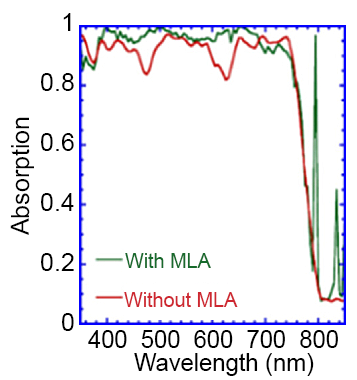}
			\caption{Simulated absorption spectrum for microlens array of period $~700 nm$ and height $800 nm$. The absorption of flat solar cell is overlaid for comparison.}
			\label{fig:peer bismwas_absorption}
		\end{subfigure}
		\caption{PPSC with microlens array on air-glass side \cite{Peer2017}. Reproduced with permission.}
	\end{figure}
	
	A. Peer \textit{et al.} \cite{Peer2017} simulated planar PPSC with a stack including a 400 $nm$ thick MAPI layer. The cell lay on top of a 700 $\mu m$ thick glass substrate. A micro lens array was then added on the top face of the glass substrate, by imprinting a polymer such as polystyrene (see Figure \ref{fig:peer bismwas_schema}). Each micro lens of the triangular lattice had a smooth profile close to truncated pyramids. For an aspect ratio period/height of the micro lenses close to 1, and a period of 700 $nm$, optimized gain of 6.3\% of the $J_{sc}$ was reported compared to the flat reference, whose EQE noticeably shown dips likely due to cavity effects in the cell (see Figure \ref{fig:peer bismwas_absorption}). Then, according to the EQE of the cell coupled to the micro lens array, the overall resulting broadband enhancement of the absorption is due at lower wavelength to AR effect, and to LT at the band edge. The optimal period of the pattern is rather large, in agreement with a low effective index of the coupled guided mode. It means that LT occurs in the mode that is hardly guided in the perovskite, but that has a nonnegligible overlap with the pattern. This overlap remains anyway limited, as revealed by the high quality factor of the resonance. Therefore, the absorption reaches almost 1, so is close to critical coupling conditions.
	
	\begin{figure}[h]
		\centering
		\begin{subfigure}[b]{0.45\textwidth}
			\centering
			\includegraphics[width=\textwidth]{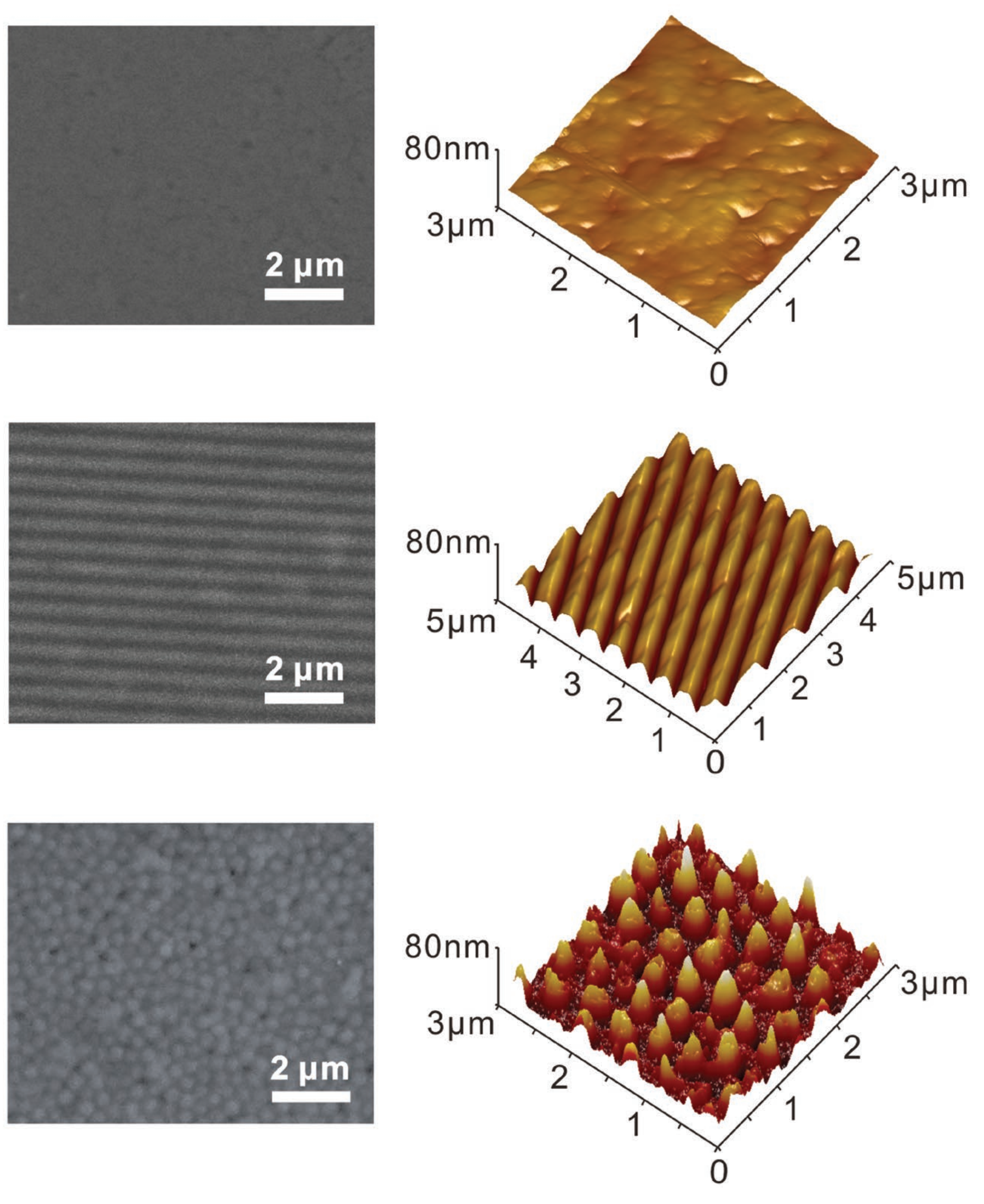}
			\caption{SEM (left)  and  AFM (right) images of  the spin-coated layer (top) of grating patterned layer (middle) and moth-eye  patterned layer (bottom).}
			\label{fig:wei 2017 adv Energy mat afm}    
		\end{subfigure}
		\hfill
		\begin{subfigure}[b]{0.45\textwidth}
			\centering
			\includegraphics[width=\textwidth]{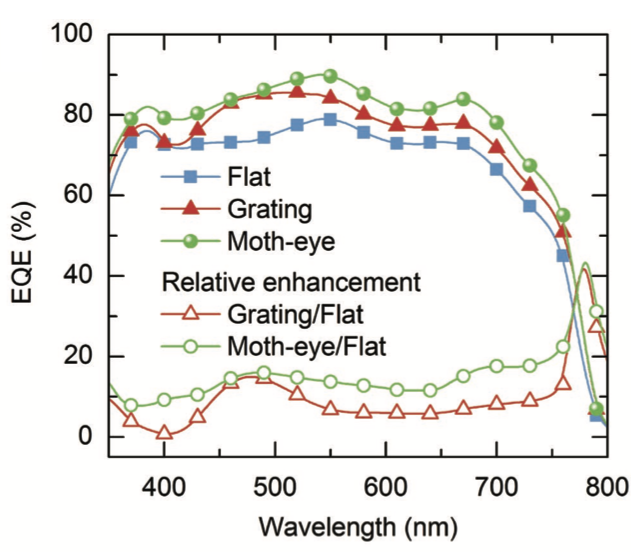}
			\caption{External quantum efficiency  (EQE) spectra of PPSC, and relative enhancement obtained by dividing the spectra of grating and moth-eye patterned devices by that of flat one.}
			\label{fig:wei 2017 adv Energy mat perf}
		\end{subfigure}
		\caption{PPSC with bio inspired PCBM HTL \cite{Wei2017}. Reproduced with permission.}
	\end{figure}
	
	J. Wei \textit{et al.} \cite{Wei2017} fabricated bioinspired back electrodes by imprinting the HTL made of PCBM before the conformal deposition of the Bphen/Ag back contact. Patterns were either a periodic sinusoidal 1D grating or a uniform 2D moth-eye structure (see Figure \ref{fig:wei 2017 adv Energy mat afm}), with a typical pitch of about 600 $nm$. It has then been experimentally shown (see Figure \ref{fig:wei 2017 adv Energy mat perf}) that the EQE of the 240 $nm$ thick MAPICl PPSC in increased by 10\% up to 40\% at the band edge, mainly thanks to the 2D moth eye pattern. The overall $J_{sc}$ increased by 14.3\%, mainly due to absorption increase, but a lower series resistance was also measured for patterned cells, which led to a slight increase of the IQE. As shown by there FDTD simulations of periodic structures, this is due light diffraction into the guided modes, rather than plasmonic effects. It can be confirmed by Fourier Analysis of the top view of the moth eye pattern that a significant part of the spatial frequencies lies in the optimal range for LT (see Figure SI-\ref{fig:Supplemental Fourier Analysis Wei}).

	\begin{figure}[!h]
		\centering
		\includegraphics[width=\linewidth]{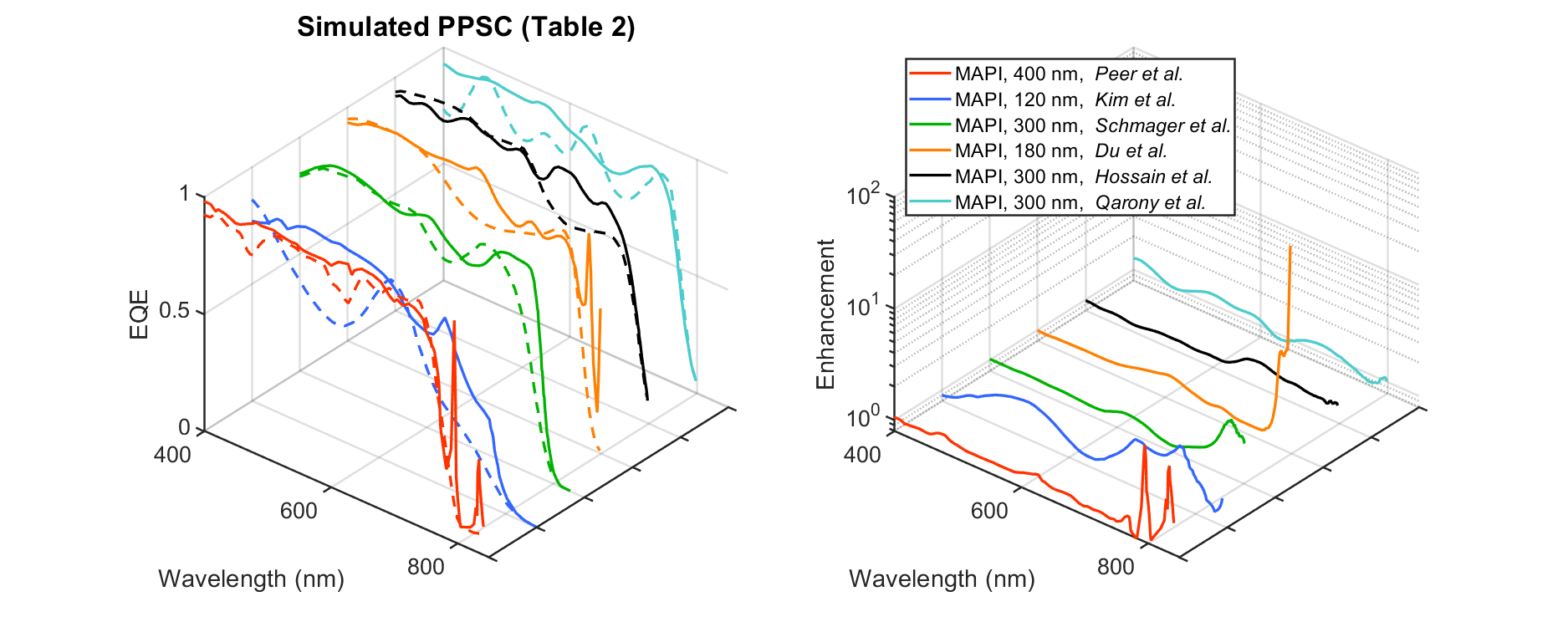}
		\caption{EQE of the patterned PPSC (solid line) and of their planar references (dashed line) (left) and EQE enhancement ((right) of reported simulated $J_{sc}$ enhancements of Table \ref{tableau pattern mapi}, mainly resulting from Light Management, for MAPI single junction PPSC, with any kind of in-plane pattern (additional layers, structuration of the perovskite or conformal perovskite) and whatever the main identified photonic concept \cite{Peer2017,Wei2017,kim_light_2021,Schmager2019,Du2016,Hossain2020,Qarony2018}}
		\label{fig:EQE_table_2}
	\end{figure}

	H. Kim \textit{et al.} \cite{kim_light_2021} simulated nanosphere arrays of TiO$_2$ conformally coated with silica on a standard MAPI PPSC stack, with a MAPI thickness limited to 100 $nm$. Provided suitable spacing and nanosphere diameter, the $J_{sc}$ could jump from 14 to 18.7 $mA.cm^{-2}$. Authors invoked the Mie scattering effect of the array to explain the enhancement, but it appears that the FEM simulation has likely been done using periodic boundary conditions, and thus that a LT effect occurs at several wavelengths, such as 705 or 790 $nm$ for those at the band edge of MAPI. This is in line with the larger EQE enhancement close to the band edge, as can be seen in Figure \ref{fig:EQE_table_2}. It remains that the overall $J_{sc}$ is anyway limited due to the unusually low thickness. \\
	
	\paragraph{Structuration of the perovskite} 
	
	\begin{figure}[!h]
		\centering
		\begin{subfigure}[b]{0.45\textwidth}
			\centering
			\includegraphics[width=\textwidth]{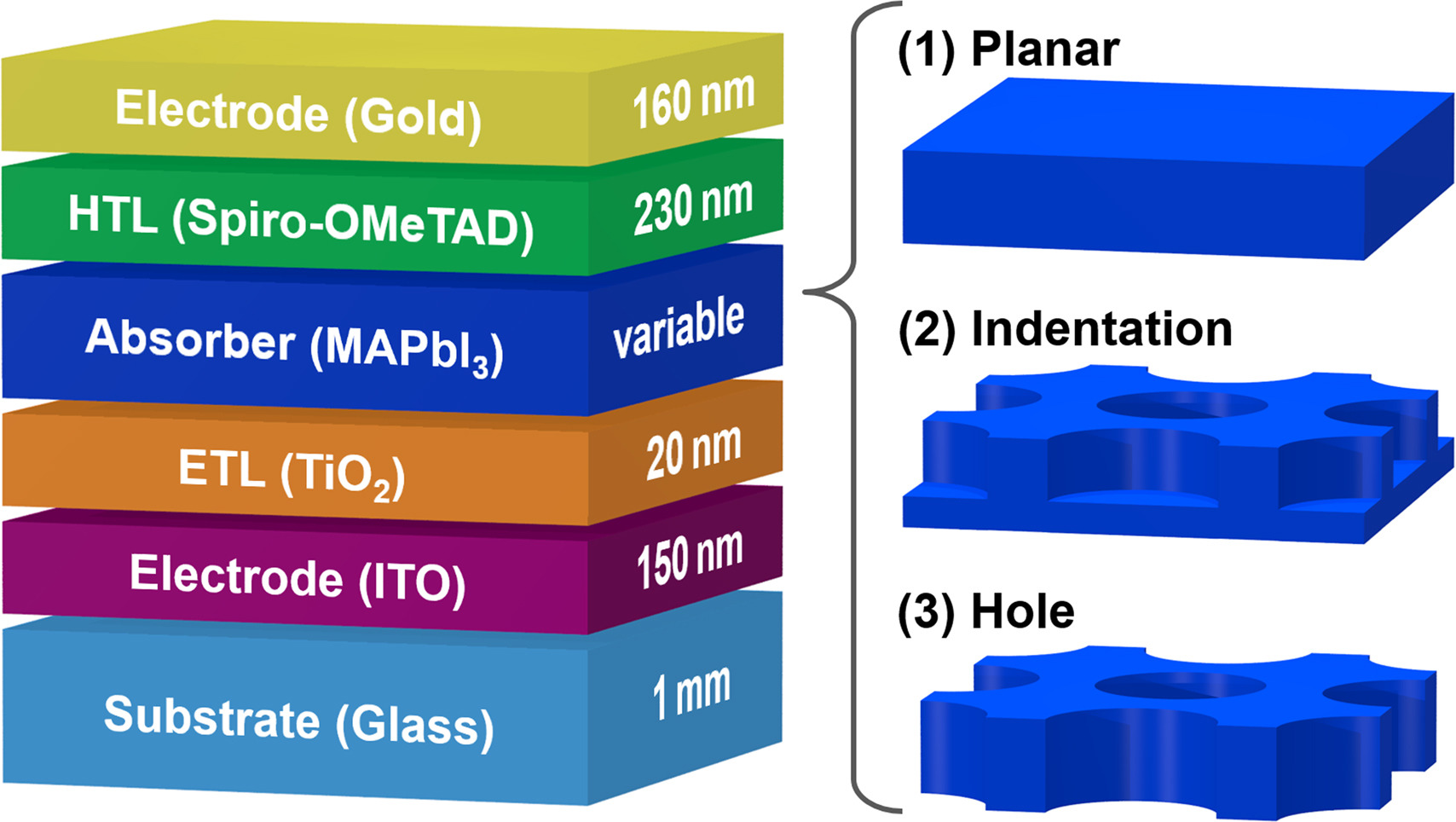}
			\caption{Layer stack of the simulated patterned PPSC, along with associated layer thicknesses. Three configurations of the active layer are simulated (1) a planar reference, (2) a cylindrical indention into the perovskite layer of variable depth; and (3) a hole geometry which corresponds to the maximum cylindrical indention.}
			\label{fig:schmager solmat schema}
		\end{subfigure}
		\hfill
		\begin{subfigure}[b]{0.45\textwidth}
			\centering
			\includegraphics[width=\textwidth]{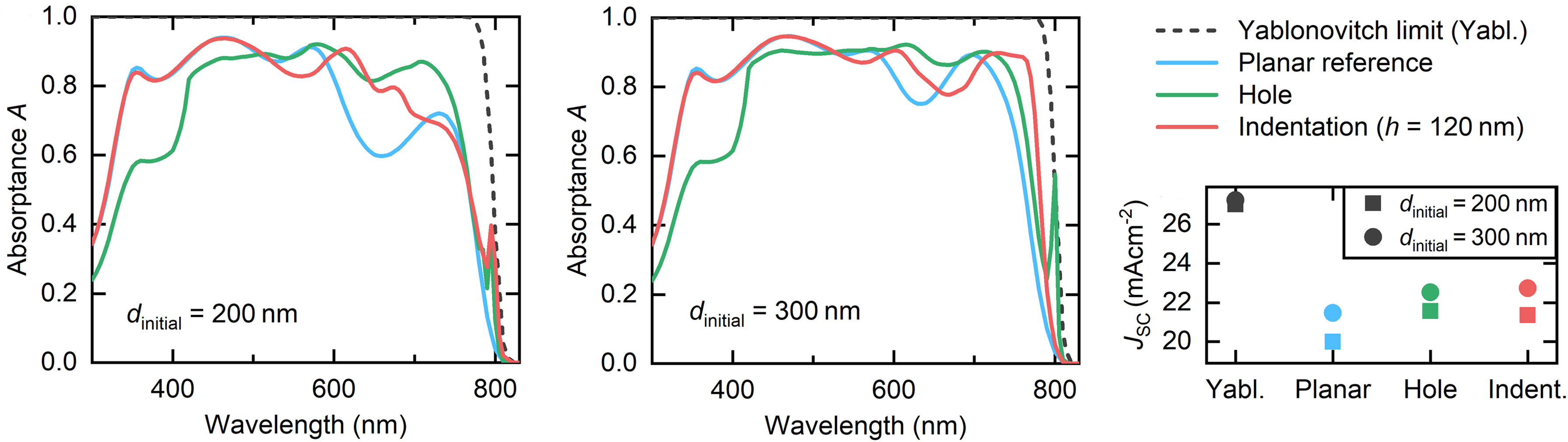}
			\caption{Absorptance in the perovskite layer of the hole pattern and the 120 $nm$ deep indentation compared to the corresponding planar reference and the theoretical Yablonovitch limit for two different initial perovskite layer thicknesses of (left) 200 $nm$ and (middle) 300 $nm$. $J_{sc}$ for all data are compared (right). The displayed nanophotonic patterns employ a geometrical $ff=0.4$ and a period of 380 $nm$.}
			\label{fig:schmager solmat simul}
		\end{subfigure}
		\caption{Various 2D in plane patterns of the MAPI layer in a PPSC \cite{Schmager2019}. Reproduced with permission}
	\end{figure}
	
	R. Schmager \textit{et al.} \cite{Schmager2019} simulated a classical MAPI PPSC (see Figure \ref{fig:schmager solmat schema}). Compared to the planar configuration, etching of the MAPI layer was envisaged, thanks to a square lattice of holes, filled with Spiro-OMeTAD, with various etching depths. To avoid any short circuit, an optimized partial etching of 120 $nm$ led to a 5.6\% increase of the $J_{sc}$, compared to the initially flat 300 $nm$ thick MAPI layer, provided an equivalent volume of perovskite. It results from a LT at the band edge, but without any sharp resonance, as can be observed on the spectral response of other patterns (see Figure \ref{fig:schmager solmat simul}).
	
	\begin{figure}[!h]
		\centering
		\includegraphics[width=\linewidth]{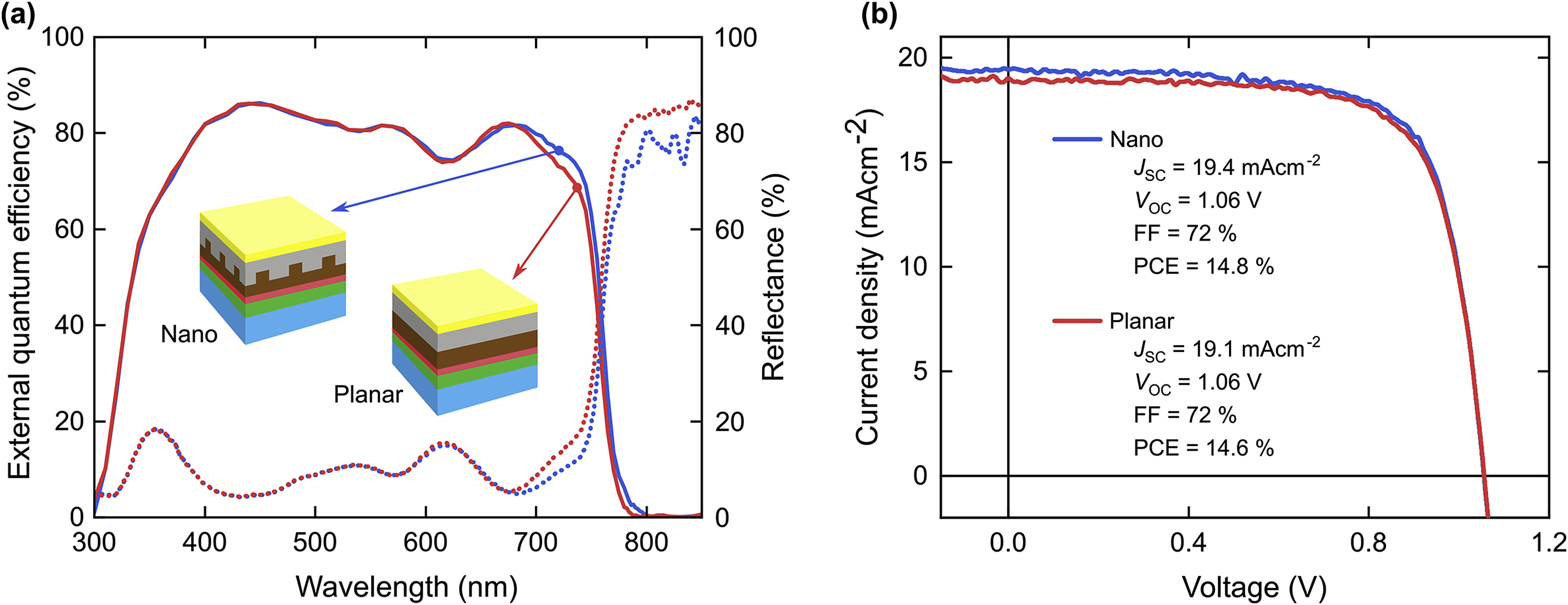}
		\caption{External quantum efficiency and reflectance spectra of the planar and nanoimprinted perovskite solar cells. The nanoimprinted perovskite layer has a period of 480 $nm$. The nanoimprinted perovskite solar cell shows enhanced absorption and current generation close to the band gap. For energies below the band gap (wavelength larger than 790 $nm$) discrete peaks are visible in the reflectance measurement. \cite{Schmager2019b}. Reproduced with permission.}
		\label{fig:schmager solmat carac}
	\end{figure}

	\begin{figure}[!h]
		\centering
		\includegraphics[width=\linewidth]{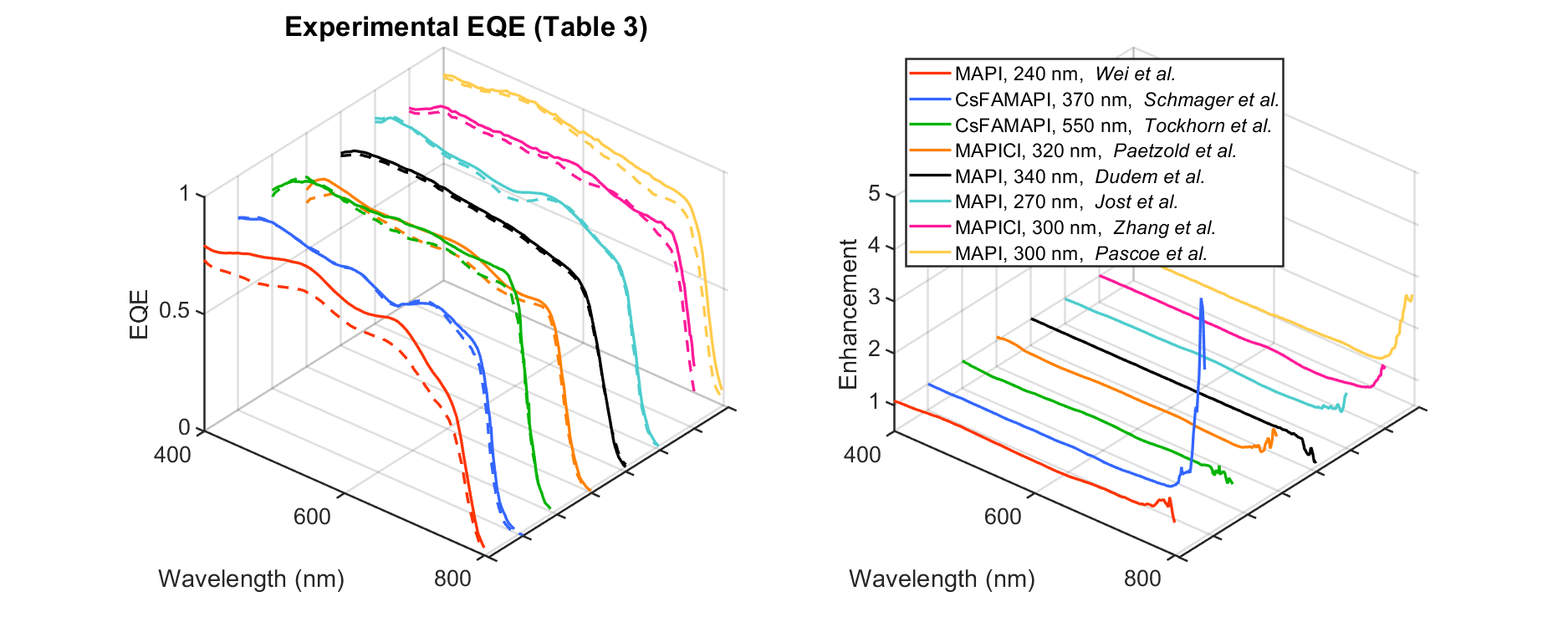}
		\caption{EQE of the patterned PPSC (solid line) and of their planar references (dashed line) (left) and  EQE enhancement ((right) of Table \ref{tableau pattern exp}, mainly resulting from Light Management, for  single junction PPSC, with any kind of in-plane pattern (additional layers, structuration of the perovskite or conformal perovskite) and whatever the main identified photonic concept \cite{Schmager2019b,Tockhorn2020,Paetzold2015,Dudem2016,Jost2017,Zhang2018b,Pascoe2016}}
		\label{fig:EQE_table_3}
	\end{figure}
	
	The same group \cite{Schmager2019b} also realized nanoimprinted PPSC made of Cs$_{0.1}$(FA$_{0.83}$MA$_{0.17}$)$_{0.9}$Pb(I$_{0.83}$Br$_{0.17}$)$_3$. Their experimental results show a relative improvement of 2\% of the PCE compared to the planar reference for the complete, gold coated solar cell. This was obtained thanks to an increase of the $J_{sc}$ from 19.1 to 19.4 $mA.cm^{-2}$ and noticeably identical $V_{oc}$ and $FF$.The EQE of the patterned cell was improved at wavelengths larger than 680 $nm$. A coupling to a quasi-guided mode is observed in the perovskite band edge (see Figure \ref{fig:schmager solmat carac}), leading to a significant EQE enhancement (see Figure \ref{fig:EQE_table_3}). \\
	
	\paragraph{Textured substrate}
	
	\begin{figure}[!h]
		\centering
		\begin{subfigure}[b]{0.45\textwidth}
			\centering
			\includegraphics[width=\textwidth]{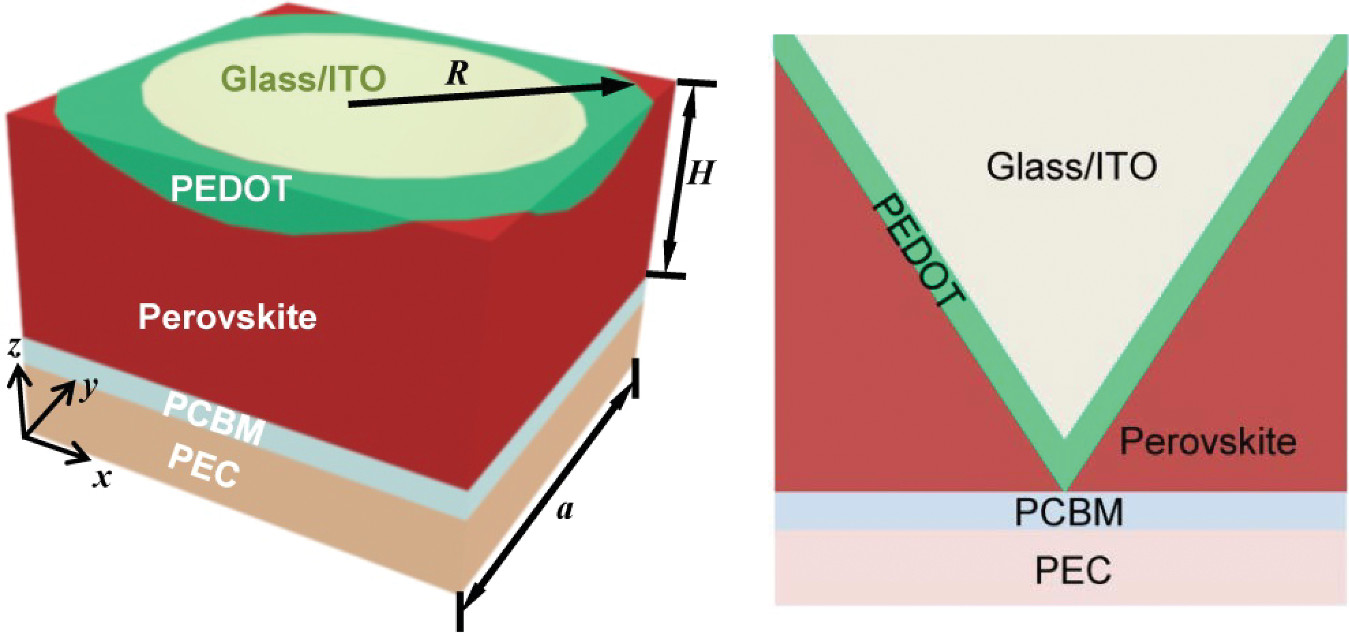}
			\caption{Three-dimensional perspective  and two-dimensional side view of the inverted vertical cone perovskite unit cell.}
			\label{fig:du john schema}
		\end{subfigure}
		\hfill
		\begin{subfigure}[b]{0.45\textwidth}
			\centering
			\includegraphics[width=\textwidth]{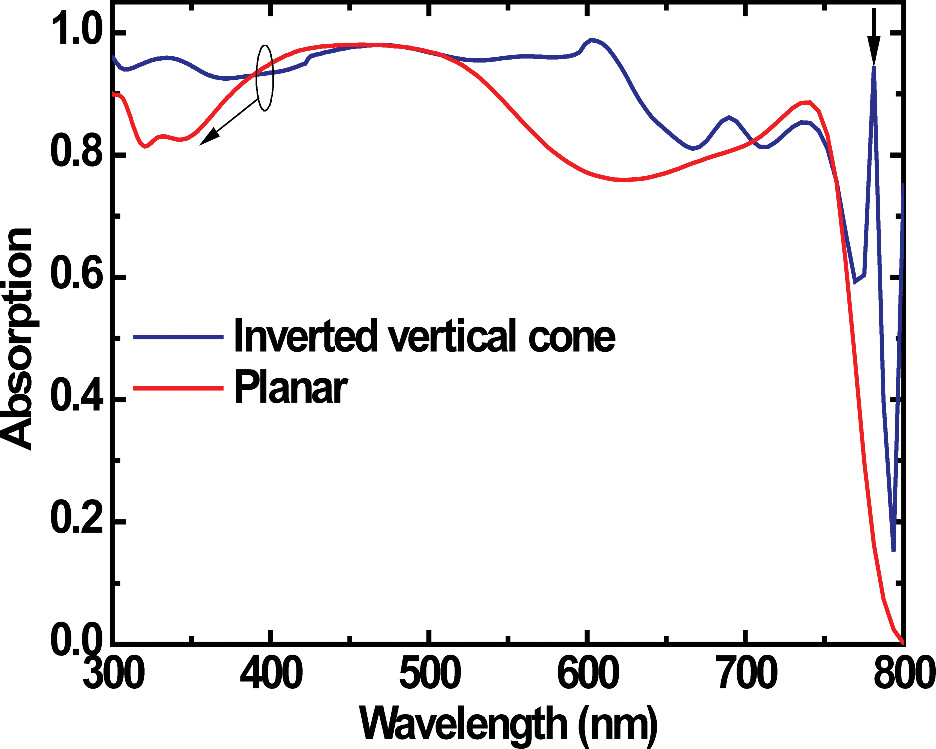}
			\caption{Simulated absorption and reflection spectra of the optimized (period $a = 600 nm$, radius $R = 380 nm$) inverted cone PC and planar PPSC. The MAPI thickness of the planar cell is 180 $nm$, corresponding to the “equivalent bulk thickness” of MAPI in the inverted-cone PC device.}
			\label{fig:du john simul}
		\end{subfigure}
		\caption{MAPI PPSC patterned with inverted cones\cite{Du2016}. Reproduced with permission.}
	\end{figure}
	
	Q. G. Du \textit{et al.} \cite{Du2016} simulated a strongly corrugated MAPI layer. It was obtained thanks to cones realized in ITO coated glass substrate with an intermediate conformal PEDOT layer lying on ITO (see Figure \ref{fig:du john schema}). Thanks to optimized period and radius, a 15-17\% improvement was expected in the PCE compared to the best flat cells of equivalent volume (see Figure \ref{fig:du john simul}). The optimal period of about 400 $nm$ fits well the required period for LT into the guided modes, given the fact that the effective index of the modes is lower than our examples, due to the strong corrugation of the MAPI filled with low index material. AR and resonant light trapping are thus simultaneously obtained.
	
	\subsubsection{Periodic structures for non resonant LM: anti-reflection and scattering}
	
	\paragraph{Superficial structuration and light management layers}
	
	M. I. Hossain \textit{et al.} \cite{Hossain2020} proposed to coat the MAPBI like perovskite, having a thickness from 100 to 400 $nm$, with zinc oxide. More precisely, a first thick, lightly doped ZnO ensured the top contact, then it could be covered by texturations patterned either as pyramids or as non-resonant metasurfaces. The period was set to 800 $nm$. Accordingly to the mechanism typically occurring using such a period, no sharp resonance enhances the absorption at the band-edge, only an AR like effect occurs. Both kinds of patterns lead to simulated equivalent $J_{sc}$ enhancements of about 3 $mA.cm^{-2}$ whatever the thickness of the perovskite between 100 and 400 $nm$, so a relative enhancement roughly from 25\% to 12\% within the same range, but only 10\% for the thickness 300 $nm$.
	
	P. Tockhorn \textit{et al.} \cite{Tockhorn2020} simulated and fabricated a 550 $nm$ thick PPSC made of a mixed cation, mixed halide Cs$_{0.05}$(FA$_{0.83}$MA$_{0.17}$)$_{0.95}$Pb(I$_{0.83}$Br$_{0.17}$) perovskite material on various periodically patterned glass substrates, in p-i-n configuration. The top side of the glass substrates was coated with low index, NaF thin film. Compared to the planar reference, patterned structures could exhibit a $J_{sc}$ up to 1 $mA/cm^2$ larger. This results from a broadband enhancement of the EQE, including at the band edge, mainly attributed to AR effect, as verified by simulations where the volume of perovskite has been kept constant. This is in line with the flat EQE enhancement, as can be seen in Figure \ref{fig:EQE_table_3} The resulting PCE reaches 19.7\%, 1\% absolute above the one of the planar reference.\\
	
	\paragraph{Structuration of the perovskite} 
	
	U. W. Paetzold \textit{et al.} \cite{Paetzold2015} proposed to pattern the front ITO electrode with a square lattice of pillars. The ETL was then corrugated, as well as the about 320 $nm$ thick MAPICl layer that planarized the stack, before the HTL and Al back contact deposition. They noticed increased absorption and EQE as the lattice period decreased, up to the smallest value envisaged, i.e. 500 $nm$. This led to an increase ofthe $J_{sc}$ by 5\%. Broadband enhancements were observed, at wavelengths shorter than the band edge of the material, as well as a limited LT at the band edge, according to the EQE enhancement plotted in Figure \ref{fig:EQE_table_3}. This is likely due to the lack of large enough spatial frequencies to couple the quasi-guided modes of the structure, given a period slightly above the optimal range. In addition, there might be a slight change in the volume of perovskite between the patterned cells compared to the flat reference.\\
	
	\paragraph{Textured substrate}
	
	W. Qarony \textit{et al.} \cite{Qarony2018} calculated the EQE for three various configurations of PPSC using the same volume of perovskite and including moth eye periodical patterns with a period of about 150 $nm$, typically leading to scattering. The fist one had only a pattern at the top air / ZnO interface (thus is more comparable to former structures), but the two others considered a conformally patterned stack, either on a patterned Al substrate, or on a patterned NiO/ITO layer. The patterned Al substrate clearly led to a lower EQE due to additional parasitic absorption. Moreover, a slightly lower $J_{sc}$ was obtained with the conformal stack on the flat Al substrate compared to the top patterned stack. Even this last case exhibits a low EQE enhancement (see Figure \ref{fig:EQE_table_3}).
	
	\subsubsection{Aperiodic patterning for LM}
	
	\paragraph{Superficial structuration and light management layers}
	B. Dudem \textit{et al.} \cite{Dudem2016} proposed a multifunctional inverted micro-structured pyramidal Polydimethylsiloxane (PDMS) AR layer for enhancing the device efficiency, through AR and self cleaning. The MAPI layer was 340 $nm$ thick. The sizes of the pyramid were in the range from 1 to 10 $\mu m$, therefore too large for efficient diffraction. Compared to flat PDMS, the $J_{sc}$ increases by 0.38 $mA.cm^{-2}$ up to 21.25 $mA.cm^{-2}$, corresponding to a limited AR effect, as confirmed by the flat EQE enhancement (see Figure \ref{fig:EQE_table_3}).
	
	Rather similarly, M. Jo\v{s}t \textit{et al.} \cite{Jost2017} fabricated a so-called light management foil on the glass substrate of a planar 270 $nm$ thick MAPI cell, which led to a limited EQE enhancement (see Figure \ref{fig:EQE_table_3}). At first, the thick glass substrate prevents from a strong overlap between any guided mode into the perovskite and the foil. Moreover, whatever the glass thickness, the lack of LT is also related to the far too low spatial frequencies resulting from the texturation. Indeed, thanks to available top view of the foil, a Fourier transform analysis reveals (see Figure SI-\ref{fig:Supplemental Fourier Analysis Jost}) that most of Fourier's components lie below 7 $\mu m^{-1}$, whereas the optimum $\beta_m$ should be in the range 18 - 34 $\mu m^{-1}$ at first order of diffraction, and even larger for other orders.
	
	On the other side of stack, H. Zhang \textit{et al.} \cite{Zhang2018b} measured and modelled the impact of the roughness of the back mirror on the back scattering of MAPICl PPSC. The reference was made of a perovskite layer with a significant roughness after crystallization, coated with a thick Spiro layer that planarizes, then a flat gold mirror. With a thinner HTL, the gold mirror replicated the roughness of the perovskite layer leading to light back scattering. The PCE increased from 19.3\% to 19.8\%, mainly thanks to the $J_{sc}$ increase from 22.7 to 23.6 $mA.cm^{2}$. It is noticeable that the obtained EQE enhancement is the largest at the band edge. This could be related to the grain size that appears to be in the range 200 - 500 $nm$, so including the optimal range for efficient LT.\\
	
	\paragraph{Structuration of the perovskite}
	
	A. R. Pascoe \textit{et al.} \cite{Pascoe2016} proposed textured MAPI at the scale of several hundreds of nanometers thanks to gas crystallization. It enhanced the EQE at wavelengths larger than 550 $nm$ (see Figure \ref{fig:EQE_table_3}), thanks to an induced so-called scattering. Accordingly, the averaged $J_{sc}$ increased from 21.3 for planar references having an about 300 $nm$ thick MAPI layer to 22.1 $mA.cm^{2}$ for patterned samples of comparable volume. As previously, the noticeably large EQE enhancement close to the band edge can be related to the typical grain size of the order of 500 $nm$.
	
	\subsubsection{Synthesis}
	In the various previously described studies reporting on $J_{sc}$ enhancements thanks to LM, it can be noticed that most of the possible architectures envisaged in the section \ref{possible_architecture} have been considered, using all the PE described in section \ref{Photonic concepts description}. Tables \ref{tableau pattern mapi} and \ref{tableau pattern exp} summarize the $J_{sc}$ enhancements reported in the various previously described studies.
	
	From Table \ref{tableau pattern mapi}, focusing on simulations of periodic patterning of MAPI based single junction PPSC, it comes out that LT can indeed significantly increase the $J_{sc}$ for very thin perovskite layers. The enhancement is more limited at a thickness of about 300 $nm$. It remains that an accurate comparison of the $J_{sc}$ enhancements is not possible, since performance enhancement strongly depends on the choice of the unpatterned reference, and especially its optimization in terms of LM as discussed previously. However, the EQE enhancements (Figure \ref{fig:EQE_table_2}) can clearly reach larger values thanks to LT than for AR.
	
	From Table \ref{tableau pattern exp}, focusing on experimental results, using various perovskite materials, the $J_{sc}$ enhancements are of the same order as the simulated ones, again with the same precautions as above. As for the simulated PPSC, the EQE enhancements (Figure \ref{fig:EQE_table_3}) also reach larger values using patterns at the scale of the wavelength in the material. Moreover, the periodic case lead to the largest enhancement.
	
	\begin{table}[h]
		\caption{Reported simulated $J_{sc}$ enhancements mainly resulting from Light Management, for MAPI single junction PPSC, with any kind of in-plane pattern (additional layers, structuration of the perovskite or conformal perovskite) and whatever the main identified PE}
		\label{tableau pattern mapi}
		\begin{tabular}{@{}lllll@{}}
			\hline
			Perov.thick. ($nm$ ) & $J_{sc}$ enhanc. (\%) & Pattern. type & Photonic concept &  Ref \\
			\hline
			400 & 6.3 & periodic add. patterned layer & Broadband AR and LT & \cite{Peer2017} \\  
			120 & 31.7 & periodic add. patterned layer & Broadband AR and LT & \cite{kim_light_2021} \\  
			300 & 5.6 & periodic struct. of the perovskite & Broadband AR and LT & \cite{Schmager2019}  \\
			180 & 17 & periodic conformal perovskite & Broadband AR and LT & \cite{Du2016} \\
			300 & 10 & periodic add. patterned layer & Broadband AR & \cite{Hossain2020}  \\
			300 & 10 & periodic conformal perovskite & Broadband AR & \cite{Qarony2018}\\     
			\hline
		\end{tabular}
	\end{table}
	
	\begin{table}[!h]
		\caption{Reported experimental $J_{sc}$ enhancements mainly resulting from Light Management, for single junction PPSC, with any kind of in-plane pattern, and whatever the main identified PE}
		\begin{tabular}{@{}lllllll@{}}
			\hline
			Material & Perov.thick. ($nm$) & $J_{sc}$ enhanc. (\%) & Pattern. type & Photonic concept &  Ref \\
			\hline
			MAPI & 240 & 14.3 & periodic add. patterned layer & Broadband AR and LT & \cite{Wei2017} \\
			CsFAMAPBI & 370 & 2 & periodic struct. of the perovskite & Broadband AR and LT & \cite{Schmager2019b}  \\    
			CsFAMAPBI & 550 & 6.3 & periodic, struct. of the ETL & Broadband AR & \cite{Tockhorn2020} \\  
			MAPICl & 320 & 5 & periodic struct. of the perovskite & Broadband AR & \cite{Paetzold2015}  \\
			MAPI & 340 & 1.8 & aperiodic add. patterned layer & Broadband AR & \cite{Dudem2016} \\
			MAPI & 270 & 4.8 & aperiodic add. patterned layer & Broadband AR & \cite{Jost2017} \\
			MAPICl & 300 & 4 & aperiodic add. patterned layer & Broadband AR & \cite{Zhang2018b} \\
			MAPI & 300 & 3 & aperiodic substrate corrugation & Broadband AR & \cite{Pascoe2016}\\
			\hline
		\end{tabular} \label{tableau pattern exp}
	\end{table}

	\subsection{LM for PR}
	
	LM for $V_{oc}$ enhancement thanks to PR in PPSC is discussed in a limited number of publications.
	
	S. Nanz \textit{et al.} \cite{Nanz2019} investigated mainly theoretically the effect of various kinds of LM strategies on the PR, for multilayer stacks that are part of PPSC, mainly without the HTL and metallic contact. They were thus able to derive an upper limit $\Delta V_{oc}$ for each case, under the assumption of pure radiative recombination. According to the summarized principle reminded previously, the $\Delta V_{oc}$ resulting from PR in a patterned multilayer was in between the values obtained with the Lambertian multilayer and the rigorously planar multilayer. Indeed, the Lambertian multilayer, as it consists in a better absorber, also radiates the luminescence, whereas, for targeted thicknesses, luminescence can be partly guided, leading to recycling. Then, the quasi-guided mode of patterned multilayer led to enhanced PR compared to the Lambertian multilayer, and also an enhanced $J_{sc}$ compared to the flat multilayer.
	However, A. Bowman \textit{et al.} \cite{Bowman2020} showed that using a more realistic model including recombination, PR was rather unlikely to occur at maximum peak power even if the cell only interacts with a limited solid angle. In this context, it thus appears more promising to increase the absorption and thus the extraction, to the detriment of recycling.
	
	\section{Simulations and Perspective}
	
	As shown in the previous section, a limited number of studies, mainly focused on $J_{sc}$ enhancements, evidenced a LM effect in PPSC. This synthesis also illustrates that due to a lack of common references, different kinds of PE can hardly be compared, so as the most promising architectures for AR and light trapping. Therefore, we propose to simulate some of the promising patterns, to be able to compare their performances using typical materials and architecture of a PPSC.
	
	\subsection{Methodology}
	
	The Rigorous Coupled Wave Analysis \cite{moharam1981rigorous} suits well for the simulation of the stacks periodically patterned under plane wave illumination. We have used the $S^4$ code \cite{LIU20122233} available in the Solcore package \cite{Alonso-Alvarez2018}. The derived $J_{sc}$ are obtained using AM1.5G spectrum \cite{Gueymard1995} provided an IQE of 1. 
	
	Optical indices used are from S. Manzoor \textit{et al.} \cite{Manzoor2018} for MAPI, from K.R. McIntosh \textit{et al.} \cite{McIntosh2009} for PMMA, and from J.M. Ball \textit{et al.} \cite{Ball2015} for all other materials: Sodalime Glass as a substrate, ITO, TiO$_2$, doped Spiro-OMeTAD (as HTL material), and gold. Thicknesses are set to 100 $nm$ for top contact, 20 $nm$ for ETL, and 300 $nm$ for Au.
	
	The ETL is supposed to be a thin TiO$_2$ layer, dense and flat, to avoid scattering, and to limit parasitic absorption compared to other envisaged organic materials.
	\subsection{Optimization of the key layers thicknesses in various planar single-junction PPSC}
	
	Let us first consider a planar PPSC on an infinitely thick substrate, illuminated through this substrate under normal incidence (see Figure \ref{fig:perspective_single_junction}). The thicknesses of the MAPI layer $th_{MAPI}$ and of the HTL $th_{HTL}$ are supposed to vary in the ranges 300 - 700 $nm$ and 200 - 400 $nm$ respectively. As can be seen in Figure SI-\ref{fig:jsc_thicknesses}, the $J_{sc}$ of such a cell increases with $th_{MAPI}$. Moreover, for a given $th_{MAPI}$, the $th_{HTL}$ has a non-negligible influence in the $J_{sc}$; e.g. for the smallest MAPI thickness of the considered range, the $J_{sc}$ can be increased by more than 0.5 $mA.cm^{-2}$, up to about 21.74 $mA.cm^{-2}$.
	
	It remains that the low $J_{sc}$ for the thinnest considered MAPI is due to lower absorption at long wavelengths, as can be seen in Figure SI- \ref{fig:absorption_spectra}, for a given thickness of HTL of 240 $nm$.
	
	Given the possible identified advantages of using a thinner MAPI layer, mainly for electrical properties, its thickness will be set to 300 $nm$ in the following.
	An AR PMMA layer is then coated on the glass substrate, which thickness is set to 1 $mm$ (see Figure \ref{fig:perspective_single_junction}), with a negligible roughness. For the chosen $th_{MAPI}$, the coupled influence of PMMA thickness, $th_{PMMA}$, and $th_{HTL}$ is studied. According to Figure SI- \ref{fig:jsc_thicknesses_pmma}, a significantly increased $J_{sc}$, up to about 21.92 $mA.cm^{-2}$, can be obtained for $th_{PMMA} = 360\ nm$ together with $th_{HTL} = 250\ nm$. This last structure will be then used as a planar but optimized reference for fair estimation of the impact of PE in the following. The corresponding spectrum is drawn in Figure \ref{fig:A_tous}.
	
	\subsection{Introduction of various 2D PC to enhance the current density}
	
	\begin{figure}[h]
		\centering
		\includegraphics[width=\linewidth]{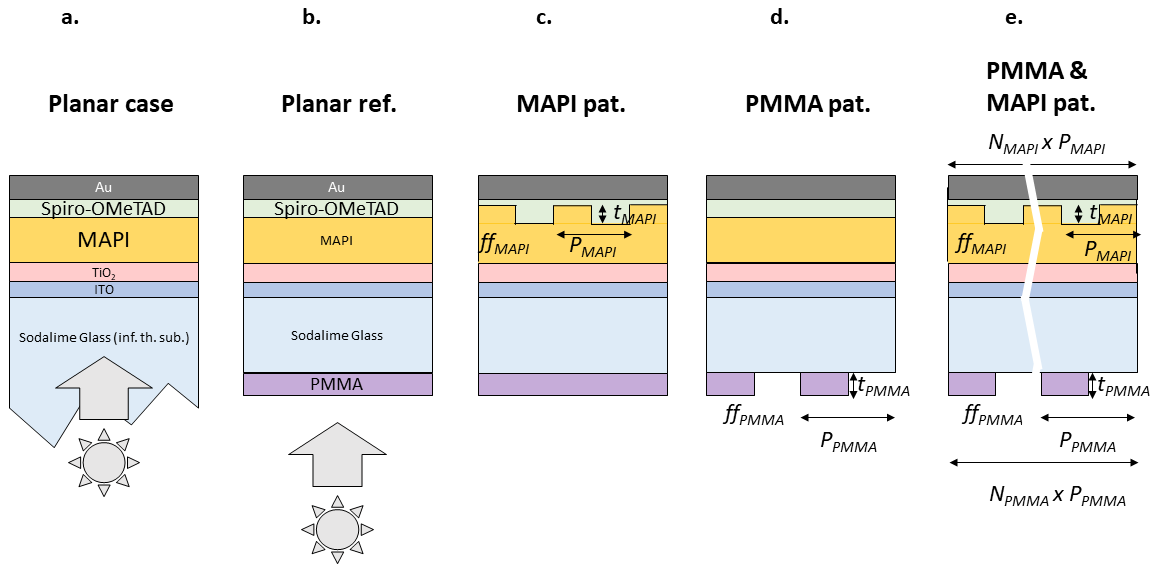}
		\caption{Various configurations of single junction PPSC simulated in the perspective. (a) planer PPSC on an infinite glass substrate, to study the effect of MAPI and Spiro-OmeTAD thicknesses (see results in Figure SI-\ref{fig:jsc_thicknesses}), (b) Planar reference having a 300 $nm$ thick MAPI layer, and various PMMA and Spiro-Ometad thicknesses to maximise absorption (see results in Figure SI-\ref{fig:jsc_thicknesses_pmma}), (c) Cross section of a 2D square lattice of cylindrical holes in MAPI, to study the effect of the corresponding PC parameters (see results in Figures SI-\ref{fig:TOTAL_absorption_Jsc_Jo_fn_period_etch_depth40_PC_perovskite} and \ref{fig:A_tous}), (d) Cross section of a 2D square lattice of cylindrical holes in PMMA, to study the effect of the corresponding PC parameters (see results in Figures SI-\ref{fig:TOTAL_absorption_Jsc_Jo_fn_period_etch_depth40_PC_pmma} and \ref{fig:A_tous}), (e) Cross section of two 2D distinct square lattices of cylindrical holes in PMMA and MAPI, to study the effect of PC parameters (see results in Figures SI-\ref{fig:scan_N_Nprime} and \ref{fig:A_tous})  }
		\label{fig:perspective_single_junction}
	\end{figure}

	As already discussed, to further increase the absorption and thus the $J_{sc}$, the most efficient strategies should be to couple the impinging light into the guided modes thanks to properly designed patterns. In the following, we envisage 2D square lattices of cylindrical patterns. Each resulting 2D PC owns typically three parameters that can be optimized: i) its period $P$, ii) its filling fraction ($ff$) which is the ratio between the hole surface and the period surface, and iii) its thickness $t$. 
	These 2D PC can be located either (see Figure \ref{fig:perspective_single_junction}):
	\begin{itemize}
		\item in the MAPI layer, made of holes in the MAPI layer filled with HTL material; for a fair comparison, the volume of MAPI material is the same as the planar PPSC, so its total thickness changes. Moreover, $t < th_{MAPI}$ to prevent from short circuits between HTL and ETL. A $t_{HTL}=250 nm$ thick slab of HTL is kept for planarization;    
		\item in the top PMMA layer, patterned in a PC of air holes, with $t = th_{HTL}$, in favor of the diffraction efficiency, given the low index of the PMMA;
		\item simultaneously at the two previous locations, but each PC has its own set of parameters.
	\end{itemize}
	
	In the following studies, all the PC parameters are scanned over realistic ranges. The step for $P$ and $t$ is 5 $nm$, whereas only 3 $ff$ have been envisaged: 0.3, 0.4 and 0.5; these appear to be the most realistic values compatible with a large area patterning at a reasonable cost.

	\begin{figure}[h]
		\centering
		\includegraphics[width=\textwidth]{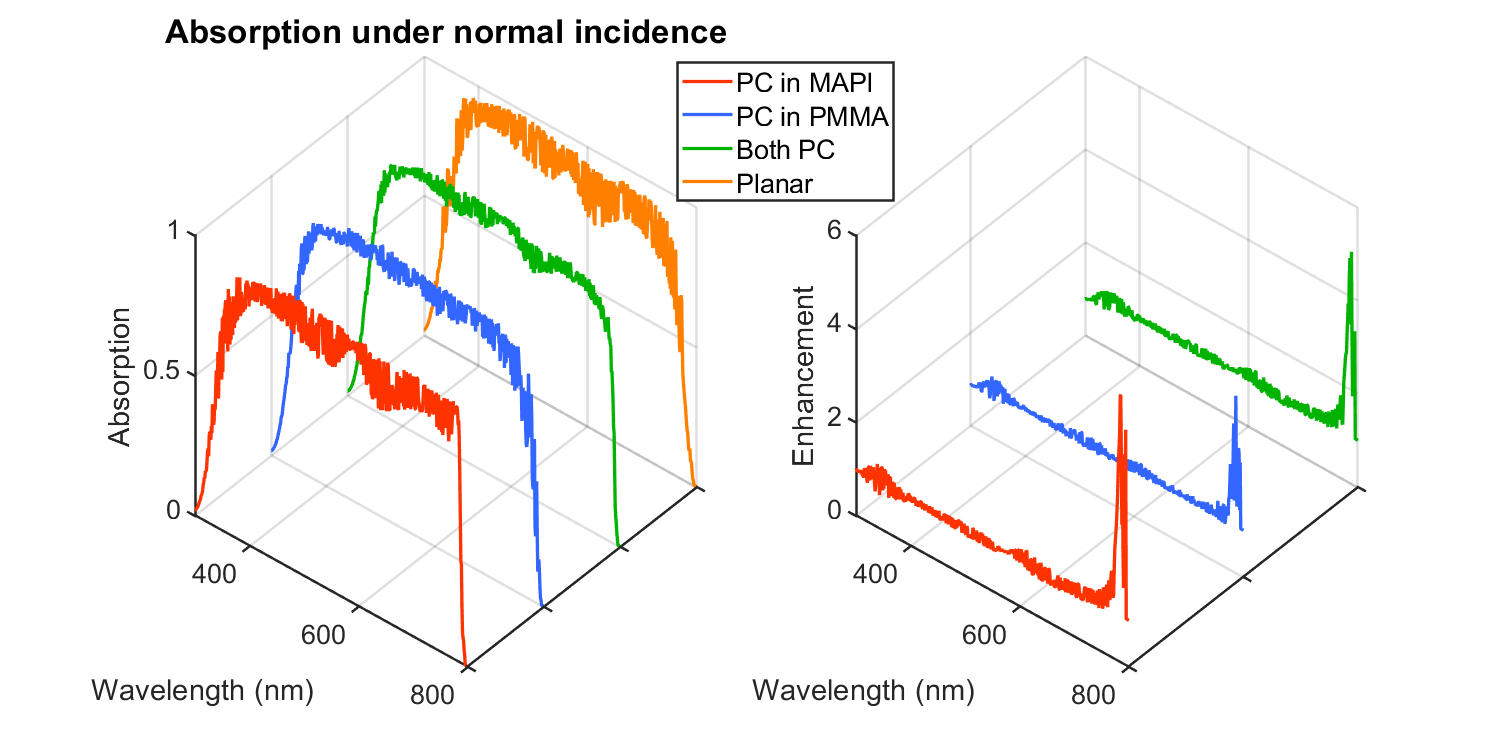}
		\caption{Simulated absorption spectra for various 1 mm thick glass substrate PPSC with maximized short circuit current density: the planar PPSC (Figure \ref{fig:perspective_single_junction} (b)), the PPSC with a 2D PC into the MAPI (Figure \ref{fig:perspective_single_junction} (c)), the PPSC with a 2D PC in the PMMA (Figure \ref{fig:perspective_single_junction} (d)), and the PPSC with both 2D PC (Figure \ref{fig:perspective_single_junction} (e)). Small amplitude, high spectral resolution Fabry Perot resonances take place in the $1 mm$ thick glass substrate.}
		\label{fig:A_tous}
	\end{figure}

	\subsubsection{Single junction: 2D PC in the perovskite layer}
	With respect to the best planar PPSC coated with PMMA, an increase of almost 1 $mA.cm^{-2}$, leading to a $J_{sc}$ about 22.88 $mA.cm^{-2}$ is obtained thanks to a PC into the MAPI (see Figure SI- \ref{fig:TOTAL_absorption_Jsc_Jo_fn_period_etch_depth40_PC_perovskite}), with $P = 400\ nm$, $ff= 0.4$ and $t= 110\ nm$. It has been checked that $ff$ of 0.3 and 0.5 lead to lower current densities than the optimal one, but still larger than the reference. The corresponding spectra in Figure \ref{fig:A_tous} reveal that the improvement is mainly due to a larger band edge absorption, so LT, as confirmed by the absorption enhancement in the same Figure.
	
	It is noticeable that other periods, smaller than 550 $nm$, can lead to more limited $J_{sc}$ enhancements. However, it has been checked that using periods around 5 and 10 $\mu m \pm 0.5$ lead to a $J_{sc}$ lower than 21.85 $mA.cm^{-2}$. It confirms that such periods, far larger than the sub-micron optimal one, do not enable an efficient diffraction and thus do not lead to any $J_{sc}$ enhancement, when compared to an optimized planar reference.  
	
	\subsubsection{Single junction: 2D PC in the PMMA covering layer}
	
	With respect to the best planar PPSC coated with PMMA, an increase of almost 0.9 $mA.cm^{-2}$, leading to a $J_{sc}$ about 22.75 $mA.cm^{-2}$, is obtained thanks to a PC in the PMMA layer. The 2D PC parameters are $P = 665\ nm$, $ff = 0.4$ and $t= 615\ nm$ (see Figure SI-\ref{fig:TOTAL_absorption_Jsc_Jo_fn_period_etch_depth40_PC_pmma}). It has been checked that $ff$ of 0.3 and 0.5 lead to lower current densities. It can be noticed on the corresponding spectra in Figure \ref{fig:A_tous} that the improvement is due to both a AR effect and limited LT at band edge absorption, since it implies low effective index guided modes that are the only able to interact, weakly, with the pattern on top of the thick substrate. The enhancement remains lower than the one induced by the 2D PC in the perovskite layer. 
	Again, it has been checked that periods far larger than the optimal one, typically around around 5 and 10 $\mu m \pm 0.5$ lead to a $J_{sc}$ lower than 22 $mA.cm^{-2}$, so to a limited enhancement, because of a reduced diffraction efficiency.
	
	\subsubsection{Single junction: Combination of the two 2D PC}
	
	Given the previous enhancements, a structure that combines the two 2D PC, one in the PMMA and the second at the MAPI/HTL interface, can be envisaged. It is noticeable that RCWA method implies that the period of such a combined architecture is a integer $N_{PMMA}$ times the period of the 2D PC in the PMMA ($P_{PMMA}$), and another integer $N_{MAPI}$ times the pitch of the 2D PC in the MAPI ($P_{MAPI}$); other PC parameters ($ff_{PMMA}$, $t_{PMMA}$ on the one hand, $ff_{MAPI}$, $t_{MAPI}$ on the other hand) can differ (see Figure \ref{fig:perspective_single_junction}).
	For the sake of illustration of a possible further $J_{sc}$ enhancement, it has been chosen to set the parameters of the 2D PC at the MAPI/HTL as for the optimized cell with a flat PMMA layer, i.e. $P_{MAPI} = 400\ nm$, $ff_{MAPI} = 0.4$, $t_{MAPI} = 110\ nm$, as well as $ff_{PMMA} = 0.5$ and $t_{PMMA} = 405\ nm$, among the thinnest most favorable values of the 2D PC in PMMA associated with planar MAPI. Then, $N_{PMMA}$ is scanned from 5 to 8 and $N_{MAPI}$ from 9 to 15, given the fact that $P_{PMMA}$ is typically larger $P_{MAPI}$ according to the previous studies. The simulated $J_{sc}$ displayed in Figure SI-\ref{fig:scan_N_Nprime} shows that a limited increase, of about 0.26 $mA.cm^{-2}$, up to 23.14 $mA.cm^{-2}$, is possible provided $P_{PMMA} = 530\ nm$. It can be seen in Figure \ref{fig:A_tous} that such $J_{sc}$ enhancement results from both LT and AR effects compared to the planar reference. If the full space of various possible PC parameters has not been scanned, and thus the previous parameters not fully optimized, the interest of such a combination is yet demonstrated.
	
	\subsection{2T tandem}
	
	\begin{figure}[h]
		\centering
		\includegraphics[scale=0.7]{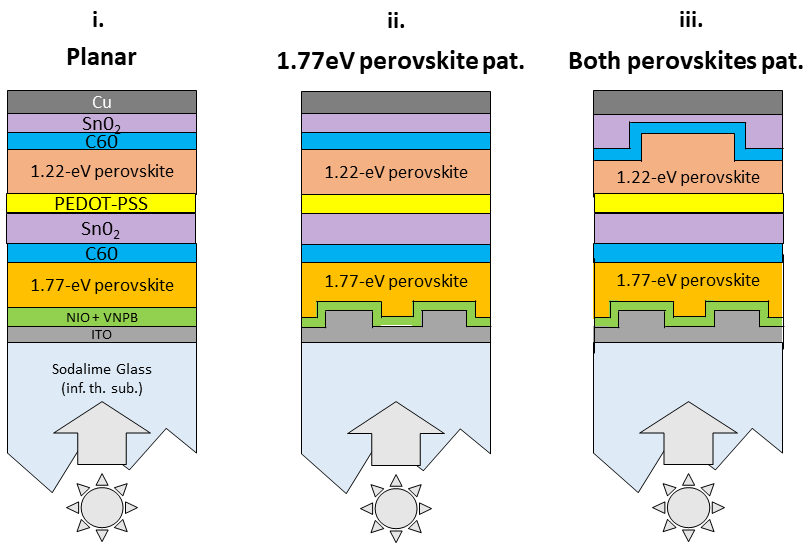}
		\caption{Various configurations of 2T tandem PPSC simulated in the perspective. (i) Planar ref on an infinite glass substrate with varying perovskite materials thicknesses, to reach and equilibrium of $J_{sc}$. (ii) Cross section of a 2D square lattice of ITO pillars in 1.77-eV perovskite, to enhance $J_{sc, 1.77eV-perovskite}$ (iii) Cross section of a 2D square lattice of ITO pillars in the 1.77-eV perovskite, and of dots of 1.22-eV perovskite in HTL. Inspired by the configuration proposed by K. Xiao \textit{et al.} \cite{xiao_all-perovskite_2020}}
		\label{fig:figures perspective tandem}
	\end{figure}
	
	A 2T tandem PPSC generally exhibits thermalization, as the one previously described in section \ref{tandem_plan}. Thus, we will consider this example as a case study for possible improvement. In the following simulations, the optical indices they provided for the considered materials are used. The planar stack (see Figure \ref{fig:figures perspective tandem}), considered as a reference, is close to the one shown in Figure \ref{fig:xiao tandem schema}. Oppositely to our previous studies, the glass substrate is again supposed to be infinitely thick to avoid the additional interferences in the substrate. Within the stack, the layer thicknesses (except perovskites) have been set to realistic values such as $t_{ITO}=100\ nm$, $t_{NiO}=t_{NVPB}= 10\  nm$ (simplified version of a mixed material ETL), $t_{C60} = 10\ nm$ for both HTL, as well as $t_{PEDOT-PSS} = 10\ nm$ and $t_{Cu}=100\ nm$. As justified later, the SnO$_2$ layer, with $t_{SnO_2}=100\ nm$, acts as an optical spacer (the 1 $nm$ thick Au layer has thus been neglect).
	
\begin{figure}[h]
		\centering
		\begin{subfigure}[b]{0.65\textwidth}
			\centering
			\includegraphics[width=\textwidth]{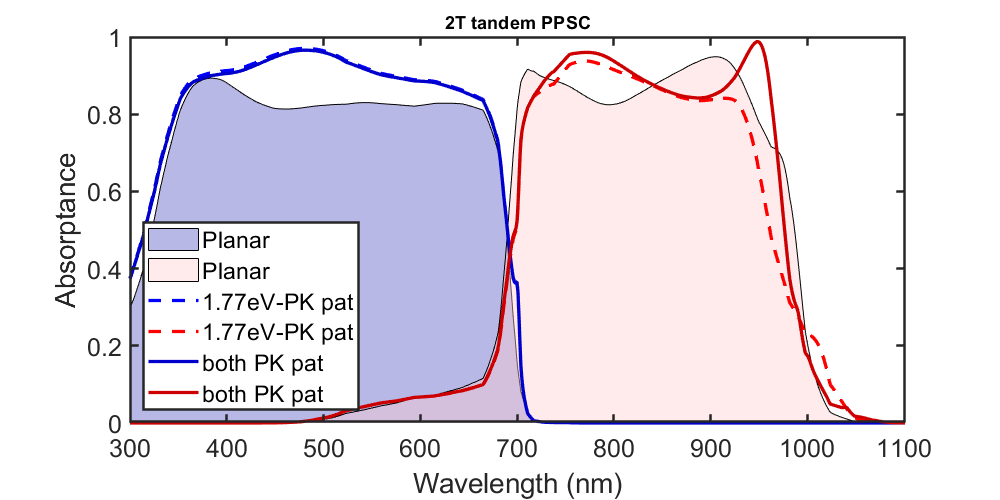}
			\caption{Simulated absorption spectra of the various envisaged 2T tandem PPSC.}
			
		\end{subfigure}
		\hfill
		\begin{subfigure}[b]{0.25\textwidth}
			\centering
			\includegraphics[width=\textwidth]{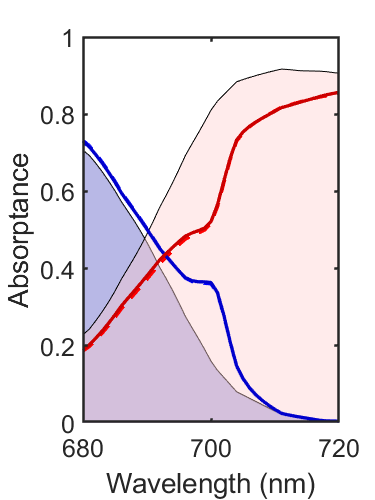}
			\caption{Zoom on LT in 1.77-eV perovskite material}
			
		\end{subfigure}
		\caption{2T tandem PPSC simulated in the perspective, Inspired by the configuration proposed by K. Xiao \textit{et al.} \cite{xiao_all-perovskite_2020}.}
		\label{fig:A_tous_tandem}
	\end{figure}

	Setting the thicknesses of both perovskite layers to $t_{1.77\ eV\ PK} = 400\ nm$ (also the maximum in the considered range) and $t_{1.22\ eV\ PK} = 880\ nm$ (in the range 800 - 1200 $nm$) leads to the highest $J_{sc}$ of 16.25 $mA.cm^{-2}$, that appears to be limited by the 1.77 $eV$ perovskite subcell.
	This value is slightly larger than the one obtained by K. Xiao \textit{et al.}. Moreover, it is obtained in our case for different perovskite thicknesses, due to a mismatch between the thicknesses of the charge transport layers we have chosen and author's choices. However, our derived absorbance spectra for both sub cells (see Figure \ref{fig:A_tous_tandem}) still exhibits a thermalization effect. Simply increasing $t_{1.77\ eV\ PK}$ could be at the expense of the charges collection. 
	
	In this frame, the possible enhancement of the $J_{sc,1.77\ eV\ PK}$ thanks to a 2D PC at the ITO / 1.77 $eV$ perovskite layer interface is studied. The 2D PC consists of a square lattice of ITO pillars, coated with conformal ETL and then with a 1.77 $eV$ perovskite layer that planarizes the corresponding subcell, while keeping an equivalent volume of perovskite as in the planar tandem cell. 
	
	A significant enhancement of $J_{sc,1.77\ eV PK}$, up to 17.92 $mA.cm^{-2}$, has been found for $P=325\ nm$, $ff=0.5$ (here defined as the ITO filling fraction) and $t=160\ nm$. 
	The reason is twofold (see spectra in Figure \ref{fig:A_tous_tandem}): a broadband AR as well as a limited LT at the band edge of the 1.77 $eV$ perovskite material that slightly reduces the thermalization. LT specifically occurs in one of the absorbing materials, to the detriment of the other one. It results from a guided mode mainly confined in the perovskite of interest, with a limited overlapping in the other perovskite, especially thanks to the rather large, 100 $nm$ SnO$_2$ optical spacer. Indeed, this layer can prevent evanescent coupling between quasi guided modes of the perovskites. Additionally, it could protect the 1.77 $eV$ perovskite material during the fabrication, but the junction between the two subcells might then also be less efficient. 
	
	Moreover, this first step leads to a strong $J_{sc}$ disequilibrium between the two sub cells, since $J_{sc,1.22\ eV\ PK}$ even slightly decreases to 15.7 $mA.cm^{-2}$, compared to the planar reference. To increase in a second step $J_{sc,1.22\ eV\ PK}$, this low-Eg perovskite layer can also be patterned rather than simply increasing its already large thickness.
	
	Starting from the last structure, a second 2D PC is introduced at the 1.22 $eV$ perovskite material / HTL interface. To target a LT at larger wavelength, close to the band edge of this perovskite, its period has to be increased. However, given the constraint induced by the boundary conditions of the simulation, we simply chose a supercell having a period twice the one of the single PC device, with one pattern in the 1.22 $eV$ perovskite material, and two patterns in the 1.77 $eV$ perovskite material (see Figure \ref{fig:figures perspective tandem}), while keeping $ff$ and thicknesses constant. Moreover, again, the volume of this perovskite is kept constant, HTL acting as a planariser, with a minimum thickness of 10 $nm$ (with a larger volume as in the planar alternative). 
	
	It appears that $J_{sc,1.77\ eV PK}$ remains unchanged, whereas $J_{sc,1.22\ eV\ PK}$ is increased up to 16.2 $mA.cm^{-2}$, reducing, but not cancelling the disequilibrium. As expected, this results from LT at the band edge in the 1.22 $eV$ perovskite material (see Figure \ref{fig:A_tous_tandem}). 
	This last result appears as a proof of concept to integrate two PC in order to induce LT, as well as possibly AR, in two different absorbing layers of a given stack, provided each one exhibits guided modes without overlapping. If not yet optimized, it is noticeable that, providing the first subcell is planarized, the two PC could practically have independent parameters, offering additional degrees of freedom to reduce disequilibrium. Given the required spacer, such a concept could be also applied to other kind of three or four terminal multijunctions cells. 
	
	\section{Conclusion - Perspectives}
	
	Light management for PV solar cells was mainly intended to increase the $J_{sc}$, simply thanks to a larger absorption. When using metal-halide perovskite materials, LM is shown to result from an interaction between the various material choices, the patterning processes and the PE. Moreover, demonstrating sole LM needs to meet some rigorous criteria to get rid of mixed electrical, material and photonic effects.
	
	The investigations presented in this review have demonstrated that the easiest $J_{sc}$ enhancement is indeed the most frequently envisaged effect, especially for single junction cells, compared to the still highly challenging $V_{oc}$ enhancement, and the immature studies of the EY due to the lack of studies at the module scale. As demonstrated by several authors, even thicknesses optimization within flat PPSC can enhance absorption thanks to Fabry Perot effects while taking care of the electrical constraints. To further increase the absorption at the perovskite band edge, an in-plane pattern, at various scales, can be introduced in one of the layers. However, patterns at the wavelength scale appeared as the most efficient. In any case, the $J_{sc}$ enhancement compared to already optimized flat reference remained limited to a few percent. As regards the photonic regimes, the broadband AR effect was the most frequently observed, but only LT was able to significantly enhance the absorption at the band edge, as confirmed by our own simulations. In addition, we confirmed that direct patterning of the perovskite leads to more efficient LT than an additional layer on top of the substrate. Finally, in 2T tandem cells, we showed that one PC in each subcell can enhance each of the $J_{sc}$ separately.
	
	Beyond, some of the strategies developed for these opaque, single junction PPSC can be adjusted for other applications of hybrid perovskites for sun light harvesting. Indeed, accurately tuned spectral absorption is required in semitransparent single junction. Moreover, the same concepts can be tweaked to tailor reflectivity spectrum for perovskite-based color printing devices \cite{Gholipour2017,Gao2018,Fan2019,Yoo2021} and engineering absorption management in full-color perovskite detectors \cite{Hossain2020b,Qarony2020}.
	
	This work is even part of far larger context. Indeed, other studies are ongoing concerning materials, which need to be more stable, and processes. In this frame, it can be noticed that record cells might not share the same encapsulation strategies as more realistic, large surface cells and modules \cite{Wang2021b}, both for aging and safety reasons \cite{wu_evolution_2021, wu_main_2021}. All perovskite tandem cells are even more challenging on the fabrication point of view \cite{Zheng2020}, but are also very promising since they combine all the potential of perovskite based cells and modules with high yields.
	
	Finally, according to the reciprocity relation between absorption and emission \cite{Rau2007}, light extraction strategies in perovskite LED \cite{Gholipour2017,Wang:17,C7NR01631J}, as well as resonances of high quality factor in perovskite-based laser \cite{Chen2016b,Pourdavoud2017,Qin2020} can use similar concepts to those derived in this work. Even LM in 2T tandem cells could be mimicked in white LEDs obtained by stacking several emitting materials. For LEDs, a high light extraction efficiency is a even more the key of good efficiency.
	
	\textit{Acknowledgement:}
	
	RCWA simulations were performed on the Newton computer cluster facilities operated by PMCS2I at Ecole Centrale de Lyon and on the CNRS/IN2P3 Computing Center in Lyon. R. M. L. acknowledges project EMIPERO (ANR-18-CE24-0016).

	\bibliography{biblio}
	
	
	
	\newpage
	
	\begin{center}
		\vspace*{\fill}
		\textbf{\large --- SUPPLEMENTAL MATERIAL ---}
		\vspace*{\fill}
	\end{center}
	
	\newpage
	\setcounter{equation}{0}
	\setcounter{figure}{0}
	\setcounter{table}{0}
	\setcounter{page}{1}
	
	\renewcommand{\theequation}{S\arabic{equation}}
	\renewcommand{\thefigure}{S\arabic{figure}}
	\renewcommand{\bibnumfmt}[1]{[S#1]}
	\renewcommand{\vec}[1]{\boldsymbol{#1}}

	\begin{figure}[ht]
		\centering
		\includegraphics[width=0.55\linewidth]{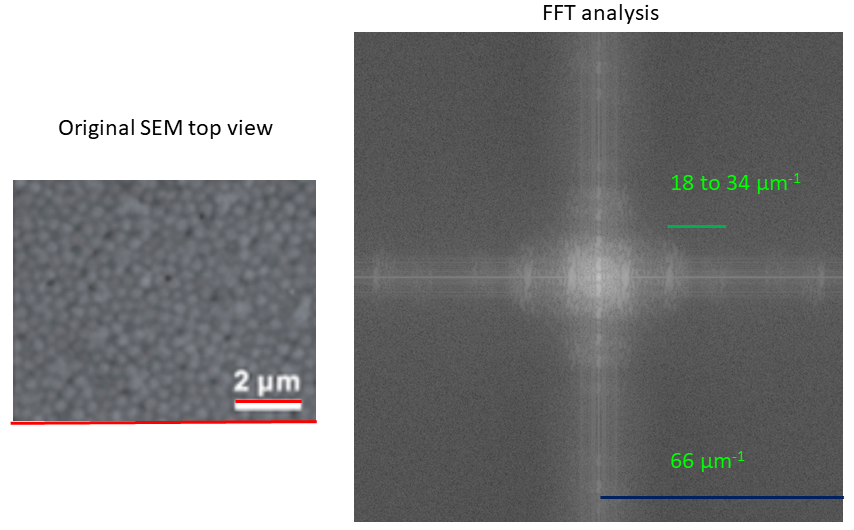}
		\caption{Fourier space analysis of the moth-eye pattern proposed by Wei \textit{et al} \cite{Wei2017}. The 615 pixels of the real space view correspond to $9.26 \mu m$; so, the resolution is $0.015 \mu m$ per pixel. Thus, the largest spatial frequency in the Fourier space is $66 \mu m^{-1}$. From equation 2, the spatial frequency range enabling the coupling of a plane wave under normal incidence into a guided mode having an effective index of two in the absorption domain of MAPI is $[18 - 34 \mu m^{-1}]$. It thus appears that some significant spatial frequencies of the pattern lie in the domain of interest, especially at large wavelengths (low spatial frequencies), partially justifying LT.}
		\label{fig:Supplemental Fourier Analysis Wei}
	\end{figure}
	\begin{figure}[ht]
		\centering
		\includegraphics[width=0.55\linewidth]{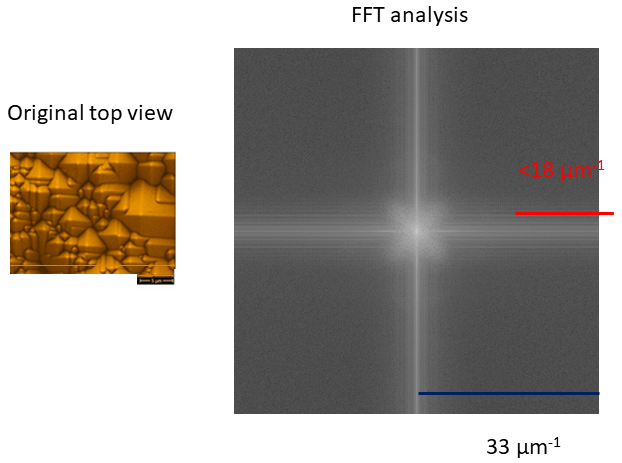}
		\caption{Fourier space analysis of the moth-eye pattern proposed by M. J\v{o}st \textit{et al} \cite{Jost2017}. The 828 pixels of the real space view correspond to $24.55 \mu m$; so, the resolution is $0.03 \mu m$ per pixel. Thus, the largest spatial frequency in the Fourier space is $33 \mu m^{-1}$. From equation 2, the spatial frequency range enabling the coupling of a plane wave under normal incidence into a guided mode having an effective index of two in the absorption domain of MAPI is $[18 - 34 \mu m^{-1}]$. Most of the significant spatial frequencies of the pattern lie outside the domain of interest. LT is thus unlikely in such a device.}
		\label{fig:Supplemental Fourier Analysis Jost}
	\end{figure}
	
	\begin{figure}[ht]
		\centering
		\includegraphics[scale=0.7]{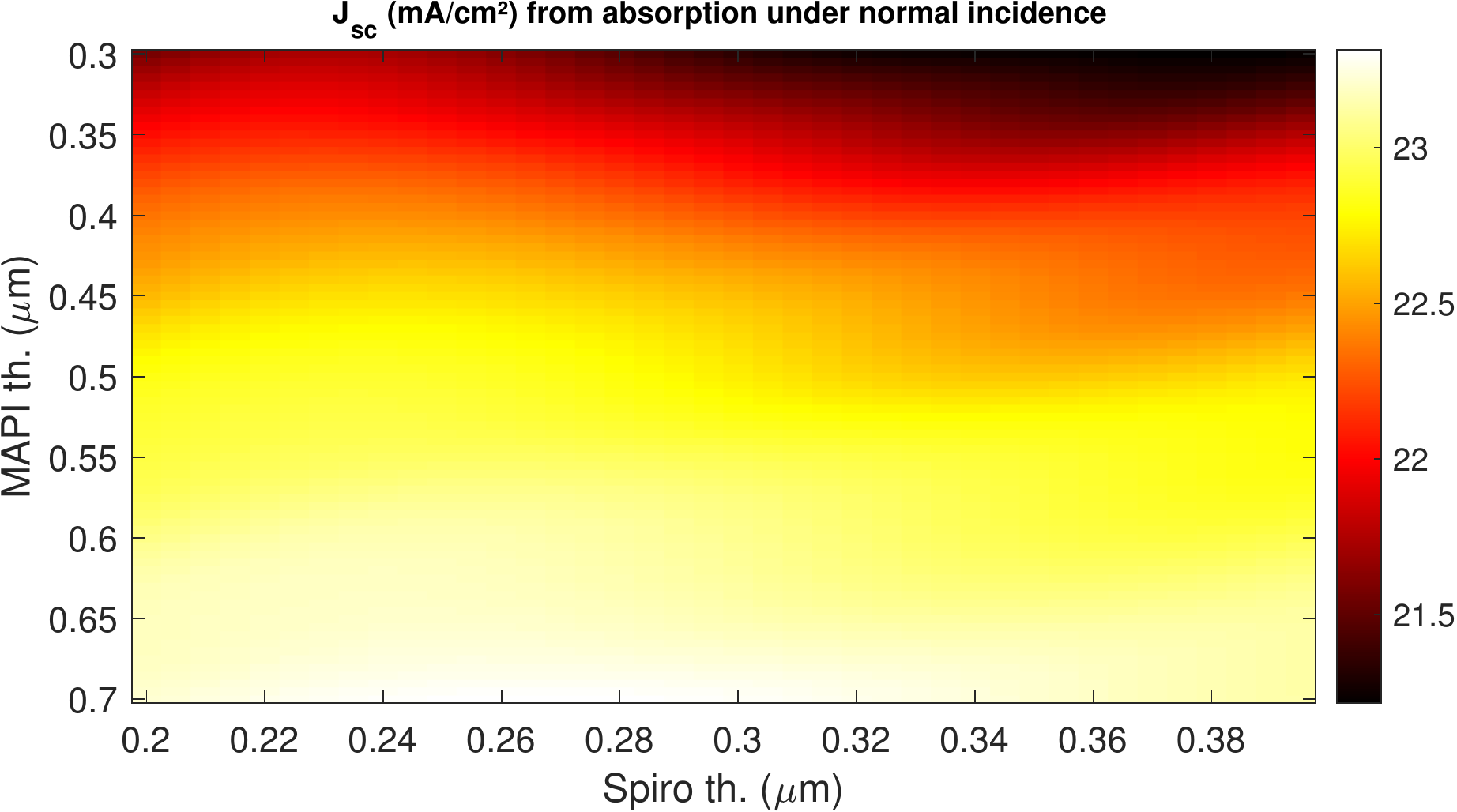}
		\caption{Simulated short circuit current density, under AM1.5G spectrum under normal incidence provided a unitary IQE for the planar MAPI PPSC as a function of HTL and perovskite thickness.}
		\label{fig:jsc_thicknesses}
	\end{figure}
	
	\begin{figure}[ht]
		\centering
		\includegraphics[scale=0.7]{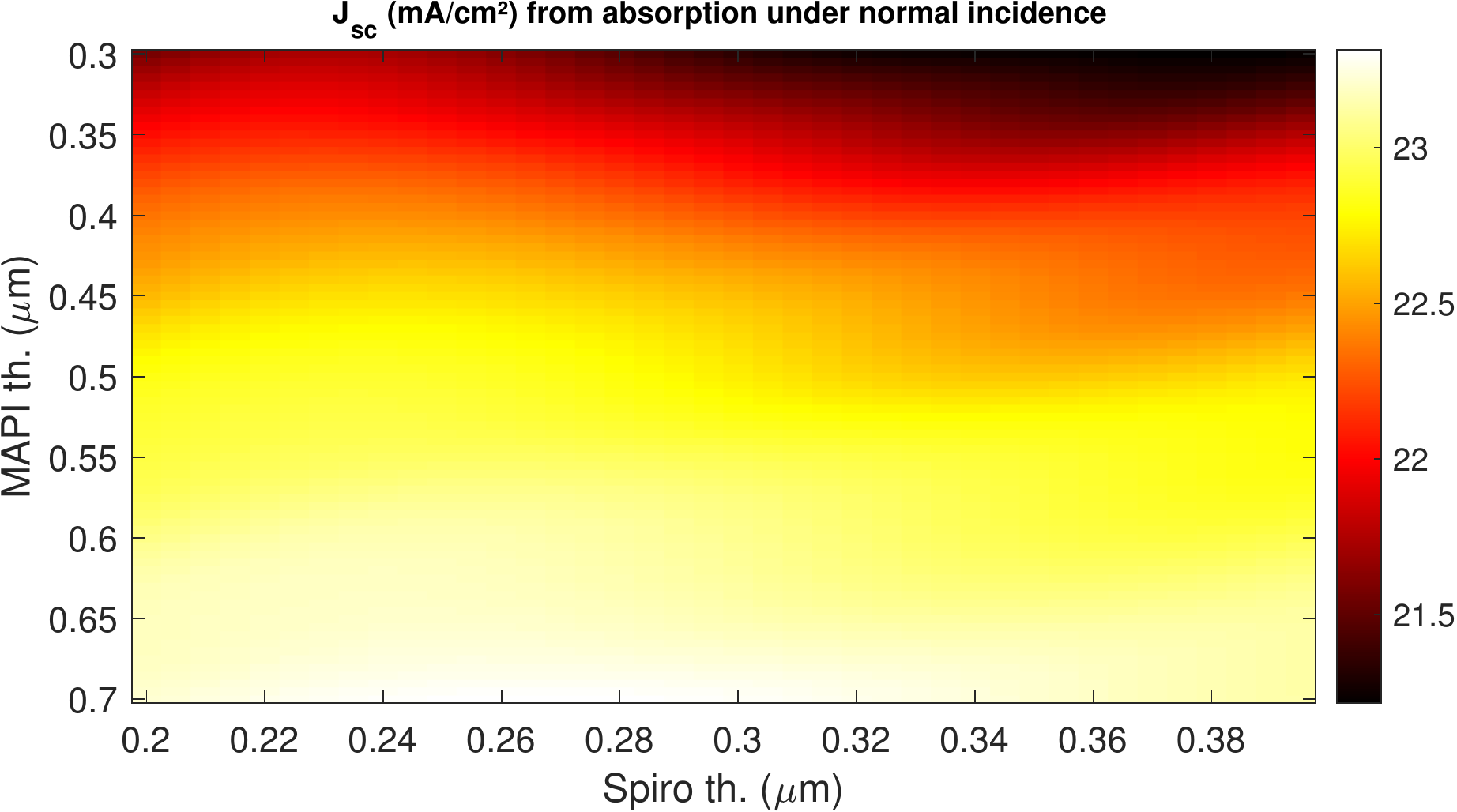}
		\caption{Simulated short circuit current density under AM1.5G spectrum under normal incidence provided a unitary IQE for the planar MAPI PPSC as a function of HTL and perovskite thickness.}
		\label{fig:jsc_thicknesses2}
	\end{figure}
	
	\begin{figure}[ht]
		\centering
		\includegraphics[scale=0.7]{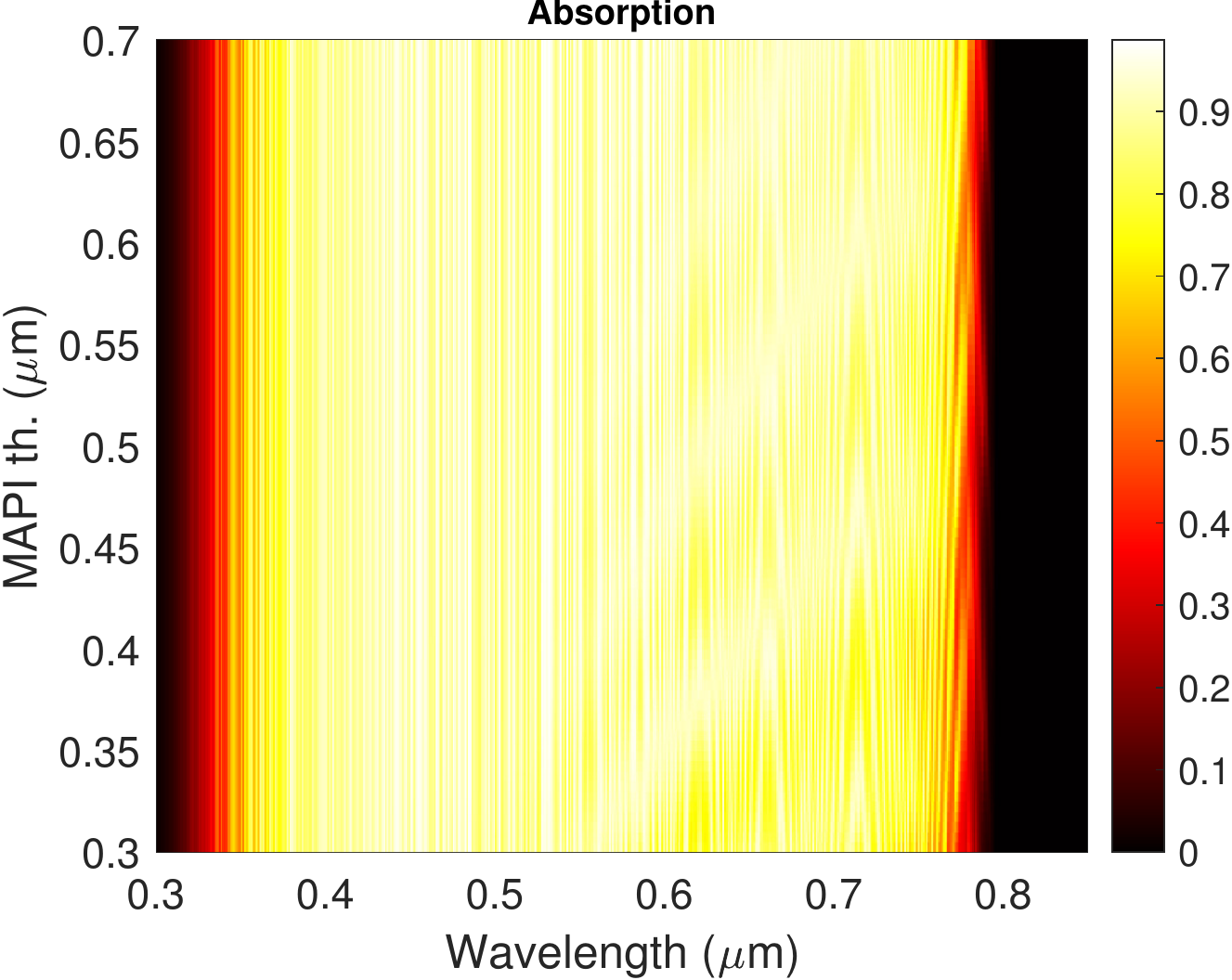}
		\caption{Simulated absorption spectra of a planar MAPI PPSC as a function of the perovskite thickness for a 0.24 $\mu m$ thick HTL.}
		\label{fig:absorption_spectra}
	\end{figure}
	
	\begin{figure}[ht]
		\centering
		\includegraphics[scale=0.7]{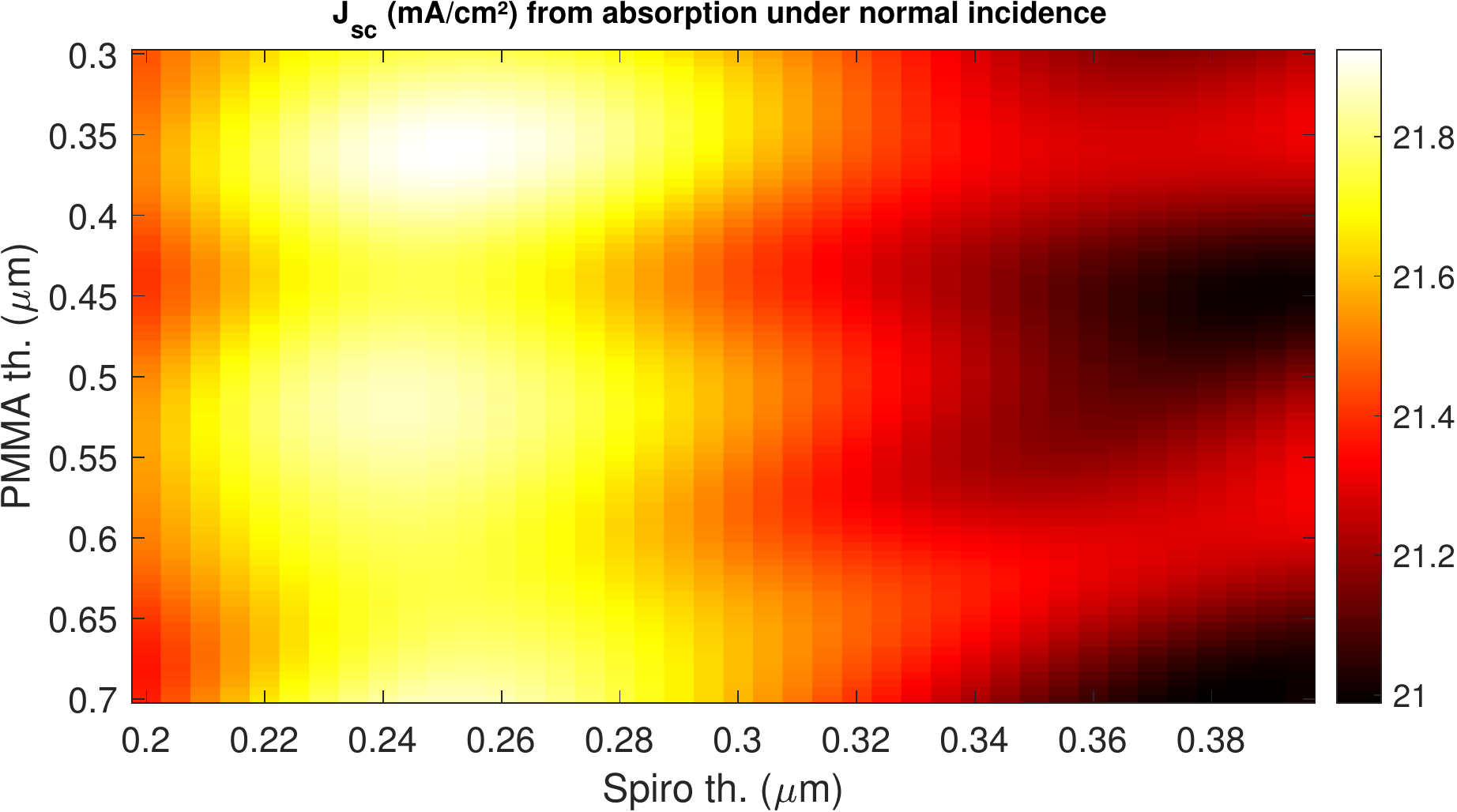}
		\caption{Simulated short circuit current density under AM1.5G spectrum under normal incidence, provided a unitary IQE of a planar 0.3 $\mu m$ thick MAPI PPSC as a function of HTL and PMMA thickness.}
		\label{fig:jsc_thicknesses_pmma}
	\end{figure}
	
	\begin{figure}[ht]
		\centering
		\includegraphics[scale=0.7]{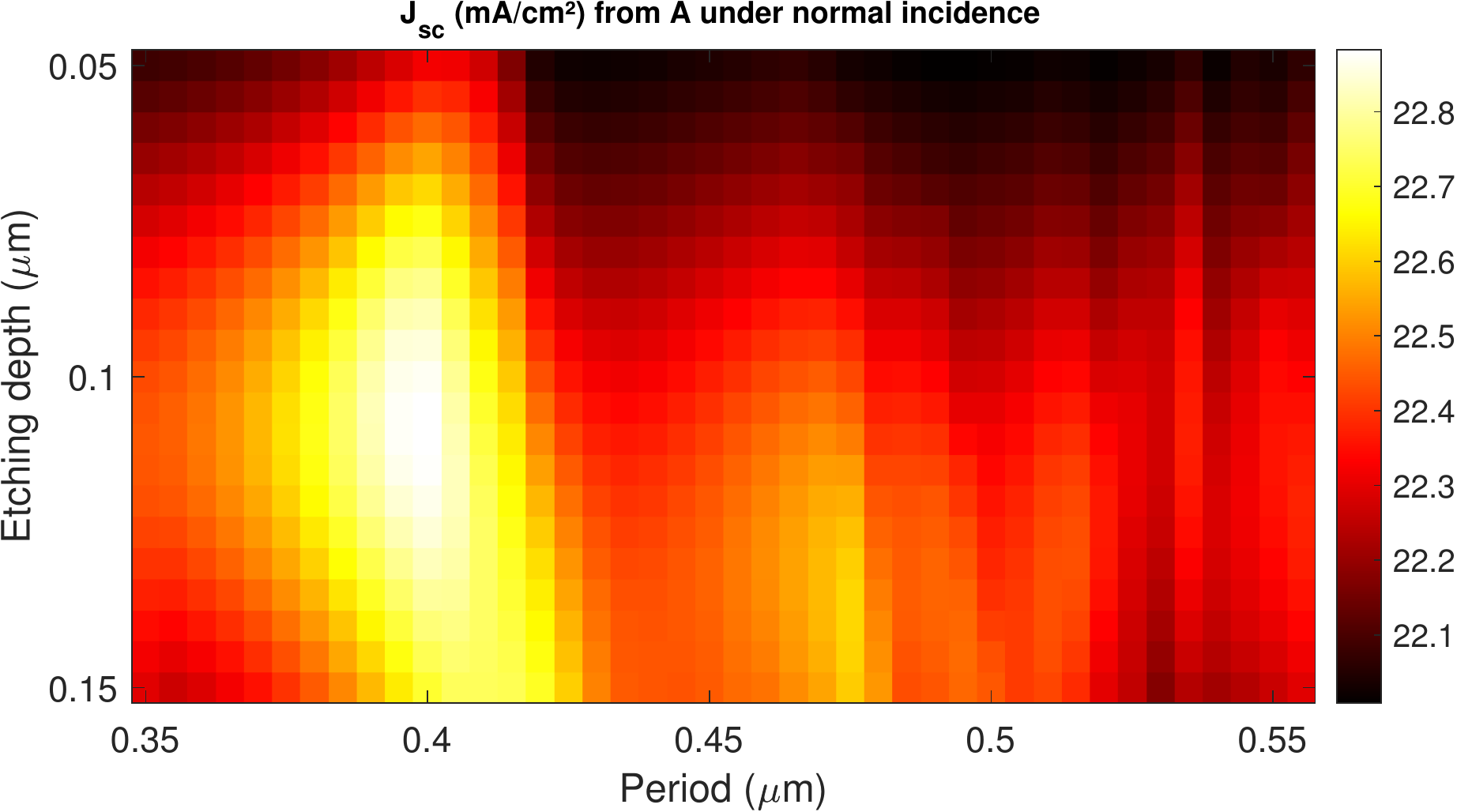}
		\caption{Simulated short circuit current density under AM1.5G spectrum under normal incidence provided a unitary IQE of a 2D MAPI PPSC with 0.25 $\mu m$ thick flat HTL, covering 0.32 $\mu m$ thick PMMA layer as a function of the hole depth and the period for $ff$ of 0.4. For all configurations, the volume of MAPI remained the same as that of the 0.3 $\mu m$ thick unpatterned layer previously considered.}
		\label{fig:TOTAL_absorption_Jsc_Jo_fn_period_etch_depth40_PC_perovskite}
	\end{figure}

	\begin{figure}[ht]
		\centering
		\includegraphics[scale=0.7]{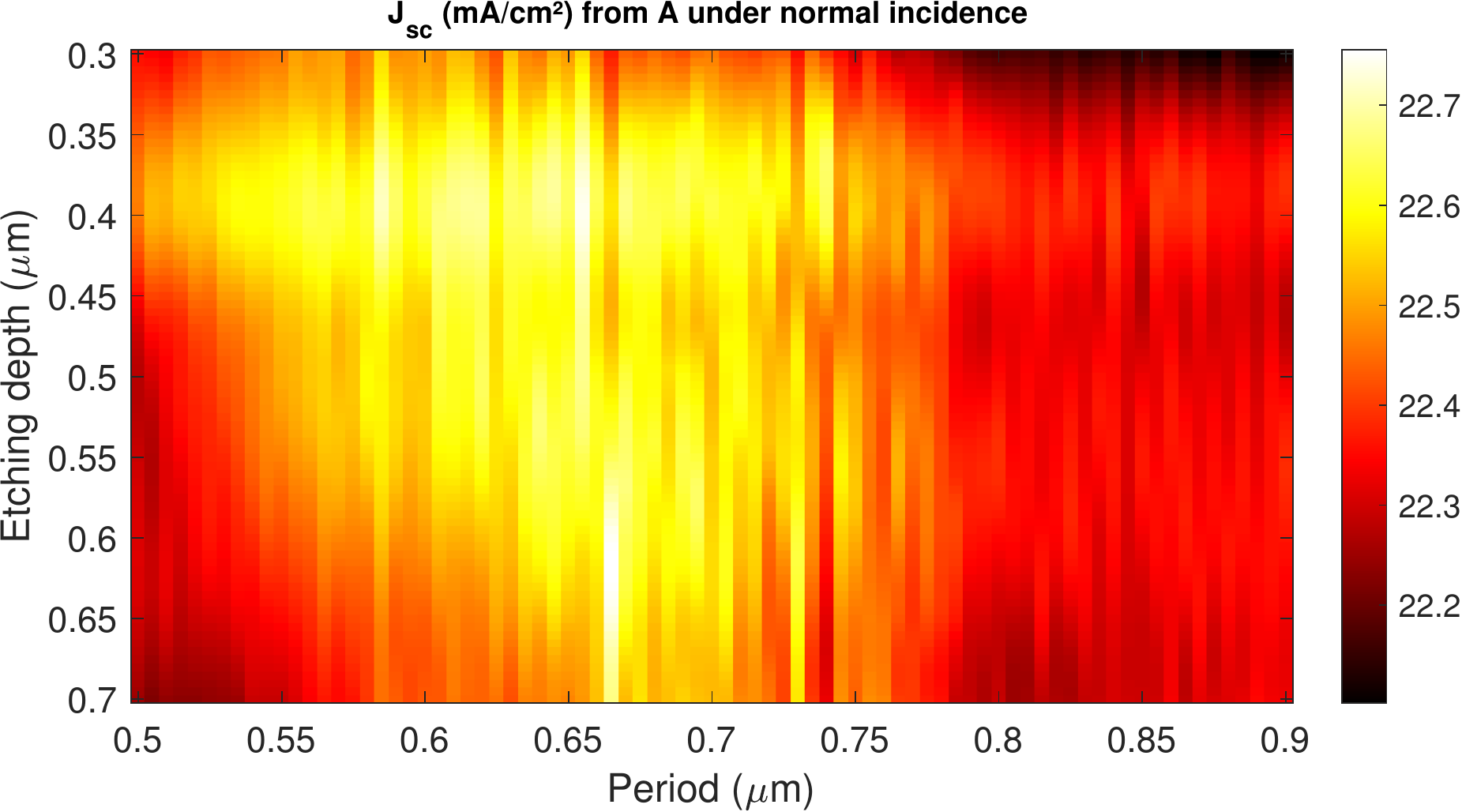}
		\caption{Simulated short circuit current density under AM1.5G spectrum under normal incidence provided a unitary IQE, of a planar 0.3 $\mu m$ thick MAPI PPSC, with a 0.25 $\mu m$ thick flat HTL coated with 2D PC PMMA layer as a function of thickness and period, for $ff$ of 0.4}
		\label{fig:TOTAL_absorption_Jsc_Jo_fn_period_etch_depth40_PC_pmma}
	\end{figure}
	
	\begin{figure}[ht]
		\centering
		\includegraphics[scale=0.7]{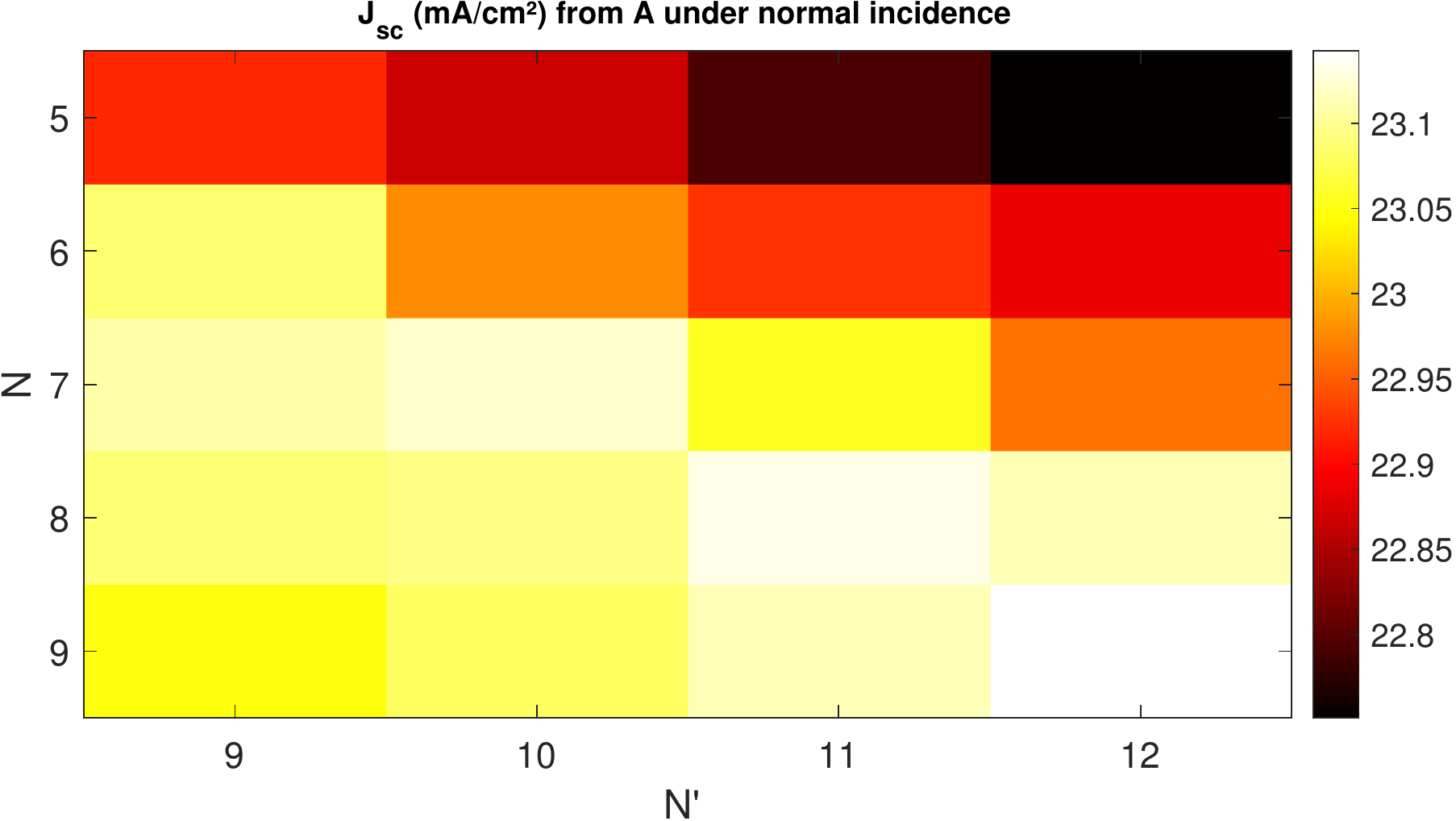}
		\caption{Simulated short circuit current density under AM1.5G spectrum under normal incidence provided a unitary IQE of a PPSC combining two 2D PCs, one in PMMA and the second at the MAPI/HTL interface. The 2D PC in MAPI has a pitch of 0.4 $\mu m$, thickness of 0.11 $\mu m$, and the $ff$ is 0.4. The $ff$ of the 2D PC in PMMA is set to 0.5, and the thickness of the PMMA layer is set to 0.405. The simulated PPSC thus has a period which is $N$ times the pitch of the 2D PC with PMMA and $N'$ times the pitch of the 2D PC with MAPI with 0.4 $\mu m$}
		\label{fig:scan_N_Nprime}
	\end{figure}
	
	\begin{figure}[ht]
		\centering
		\includegraphics[scale=0.7]{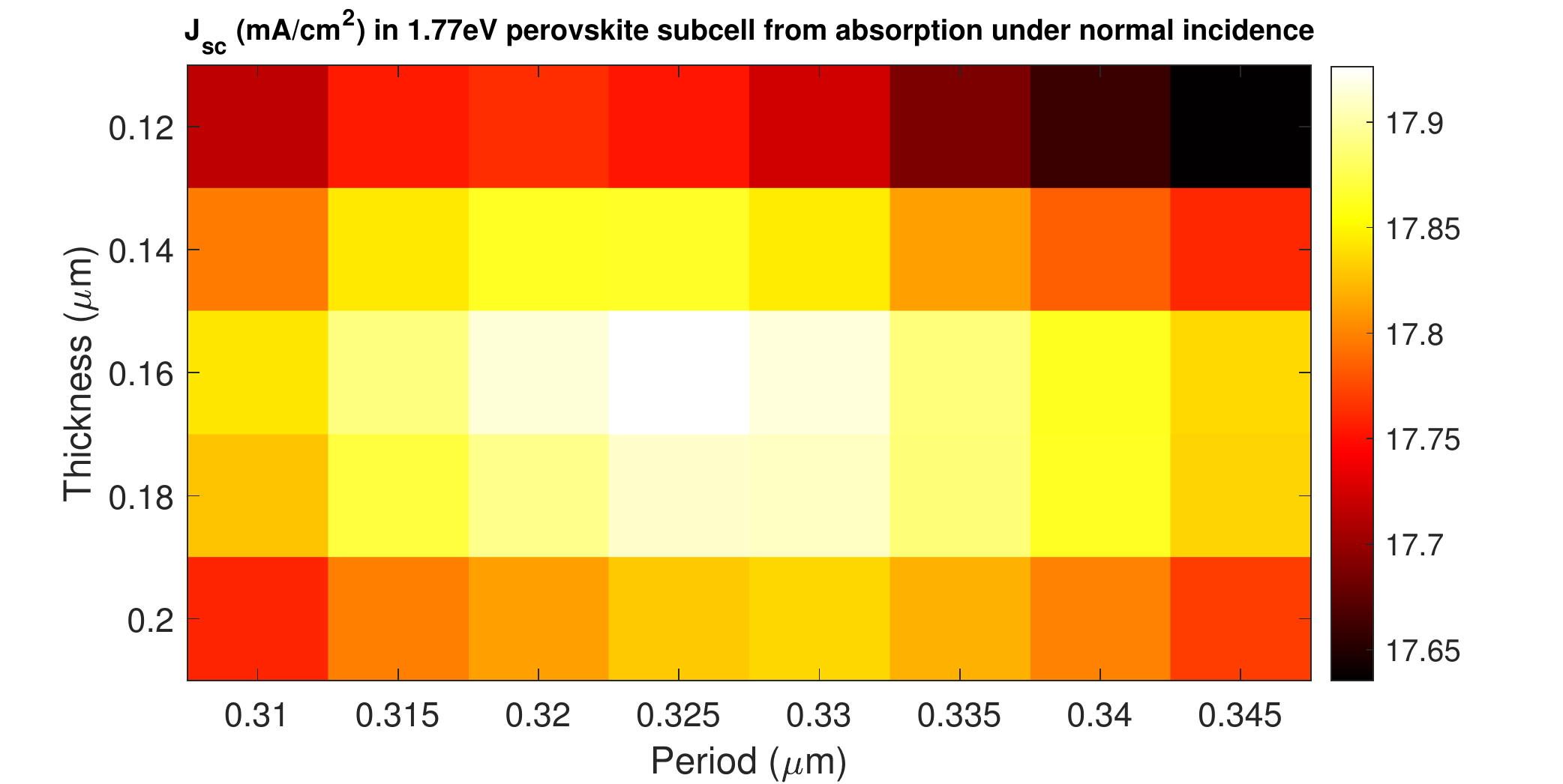}
		\caption{Simulated short circuit current density, under AM1.5G spectrum under normal incidence provided a unitary IQE, of a 2T tandem PPSC combined with a 2D PC in the large bandgap subcell as a function of thickness and period, provided $ff=0.5$.}
		\label{fig:optimisation_jsc_pk1-eps-converted-to.pdf}
	\end{figure}

\end{document}